\documentclass[sigconf]{acmart}
\usepackage{xcolor}

\newtoggle{revisions}

\iftoggle{revisions} {
\newcommand{\rev}[1]{{\color{blue}{#1}\normalfont}}
}{
  \newcommand {\rev}[1]{#1} 
 }
 
\AtBeginDocument{%
  \providecommand\BibTeX{{%
    \normalfont B\kern-0.5em{\scshape i\kern-0.25em b}\kern-0.8em\TeX}}}

\setcopyright{acmcopyright}
\copyrightyear{2024}
\acmYear{2024}
\acmDOI{10.1145/3613904.3642502}

\acmConference[CHI '24]{Proceedings of the CHI Conference on Human Factors in Computing Systems}{May 11--16, 2024}{Honolulu, HI, USA}
\acmBooktitle{Proceedings of the CHI Conference on Human Factors in Computing Systems (CHI '24), May 11--16, 2024, Honolulu, HI, USA}
\acmPrice{15.00}
\acmISBN{979-8-4007-0330-0/24/05}

\begin{document}

\title{Practice-informed Patterns for Organising Large Groups in Distributed Mixed Reality Collaboration}
\author{Emily Wong}
\affiliation{%
  \institution{The University of Melbourne}
  \streetaddress{Parkville}
  \city{Melbourne}
  \state{VIC}
  \country{Australia}
  \postcode{3056}
  }
\email{wonge3@student.unimelb.edu.au}

\author{Juan Sánchez Esquivel}
\affiliation{%
 \institution{Aarhus University}
 \city{Aarhus}
 \country{Denmark}
 }
\email{juan@cs.au.dk}

\author{Germán Leiva}
\affiliation{%
  \institution{Aarhus University}
  \city{Aarhus}
  \country{Denmark}
  }
\email{leiva@cc.au.dk}

\author{Jens Emil Gr{\o}nb{\ae}k}
\affiliation{%
  \institution{The University of Melbourne}
  \streetaddress{Parkville}  
  \city{Melbourne}
  \state{VIC}
  \country{Australia}
  }
\affiliation{%
  \institution{Aarhus University}  
  \city{Aarhus}
  \country{Denmark}
}  
\email{jensemil@cs.au.dk}

\author{Eduardo Velloso}
\affiliation{%
  \institution{The University of Melbourne}
  \streetaddress{Parkville}
  \city{Melbourne}
  \state{VIC}
  \country{Australia}
  }
\email{eduardo.velloso@unimelb.edu.au}

\renewcommand{\shortauthors}{Wong, et al.}

\begin{abstract}
Collaborating across dissimilar, distributed spaces presents numerous challenges for computer-aided spatial communication. Mixed reality (MR) can blend selected surfaces, allowing collaborators to work in blended f-formations (facing formations), even when their workstations are physically misaligned. Since collaboration often involves more than just participant pairs, this research examines how we might scale MR experiences for large-group collaboration. To do so, this study recruited collaboration designers (CDs) to evaluate and reimagine MR for large-scale collaboration. These CDs were engaged in a four-part user study that involved a technology probe, a semi-structured interview, a speculative low-fidelity prototyping activity and a validation session. The outcomes of this paper contribute (1) a set of collaboration design principles to inspire future computer-supported collaborative work, (2) eight collaboration patterns for blended f-formations and collaboration at scale and (3) theoretical implications for f-formations and space-place relationships. As a result, this work creates a blueprint for scaling collaboration across distributed spaces. 
\end{abstract}



\begin{teaserfigure}
  \includegraphics[width=\textwidth]{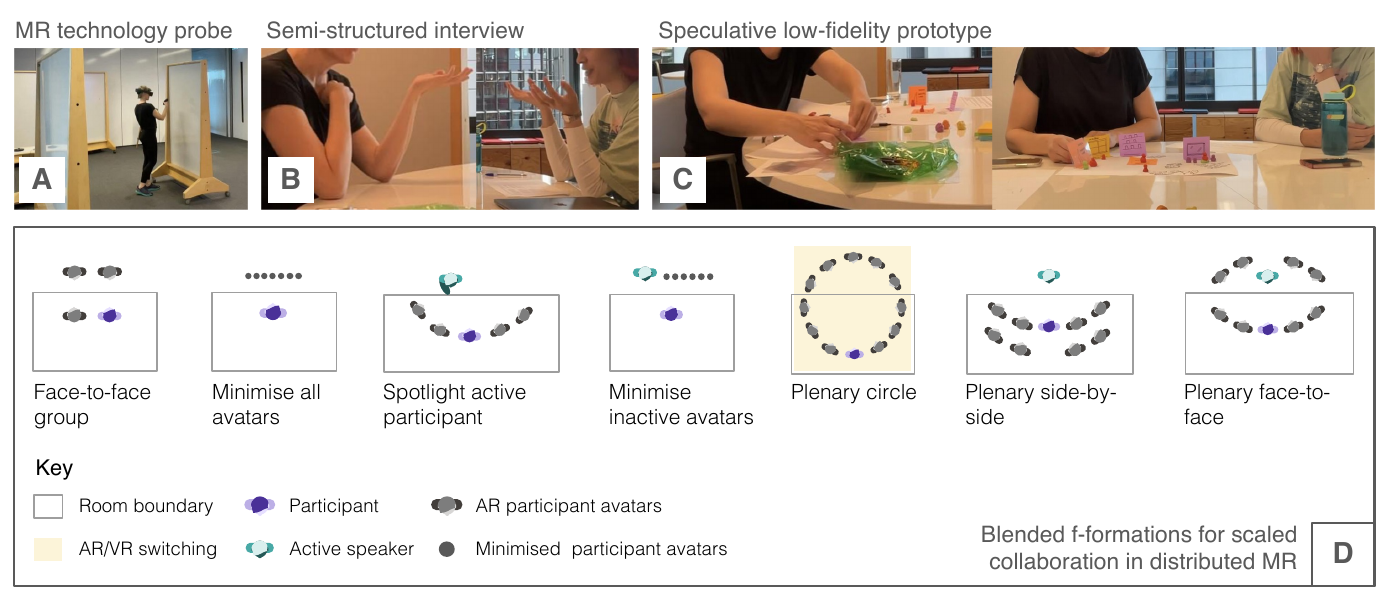}
  \caption{The user study engaged collaboration designers (CDs) in (a) a MR technology probe, (b) semi-structured interview about how they design digital and physical collaborative spaces and (c) a speculative low-fidelity prototyping activity. Working with CDs, who are experienced in large-scale collaboration, helped us to develop collaboration design principles as well as (d) blended f-formations for scaled collaboration in distributed MR.}
  \Description{xx}
  \label{fig:teaser}
\end{teaserfigure}

\received{14 September 2023}
\received[revised]{12 December 2023}
\received[accepted]{19 January 2024}

\maketitle

\section{Introduction}
Space architecture is central to the design of collaborative environments. It plays an essential role in communication, allowing co-located collaborators to relay information non-verbally through embodied interactions inside the designed space.  For example, ``when you walk into a lecture hall at a university... you still know who is the professor and who the students are'', the professor being at the front of the hall \cite[pg. 217]{buxton2009mediaspace}. People make inferences about how others position their bodies in space to make sense of the world non-verbally \cite{hall1966hidden}. This shared spatial literacy helps people self-organise and decode unspoken information quickly \cite{lawson2007language}. Understanding spatial communication supports students to quickly know when and where they should sit and if they should take notes. Social meaning is communicated thanks to a shared expectation of how people interact with each other in different spaces, such as a lecture theatre. 

Distributed collaboration, which we define to include hybrid and remote work, disrupts the way space is used to communicate social meaning. When collaborators use digital tools for distributed communication, they often do not occupy the same physical space \cite{harrison1996re, gaver1992affordances}. Depending on the digital technology used, collaborators might occupy multiple physical and digital spaces embedded within each other \cite{dourish2011divining, harrison1996re}. In teleconferencing tools such as Microsoft Teams or Zoom, each participant inhabits their own physical space while viewing a digital interface that acts as shared `space' for remote collaboration \cite{harrison1996re}.  This reorganisation of how space is co-inhabited and represented disrupts how people organise themselves spatially to communicate social meaning. Returning to our example, when a student enters a Zoom call, there is no obvious professor standing at the front of the room.  Instead, the student is greeted with many identical rectangular boxes, which lack the physical orientation that would normally indicate who the professor is.

Mixed reality (MR) helps to revitalise spatial communication during distributed collaboration. We define MR as encompassing augmented reality (AR), virtual reality (VR)~\cite{milgram1994taxonomy} and enabled by video passthrough head-mounted displays. MR technology can bring the distributed collaborator's inhabited space into the interactive environment. Blending physical objects with the interactive environment allows natural embodied communication, such as pointing to an object or turning toward a focal point. Approaches to physical space alignment include warping the avatars' positions and gestures  \cite{yoon2021full, wang2022avatarmotion, jo2015spacetimeAvatar}, digitally mapping and matching the distributed spaces \cite{johnson2023unmapped, Irlitti2023volumetric} or aligning selected objects, such as a whiteboard or desk, across the different environments \citep{gronbaek2023partially}. This alignment with the physical space, aided by MR technology, reconstructs spatial communication cues. Participants in these systems can physically turn toward the main speaker, for example a professor, to non-verbally communicate their attention. 

While MR helps to reconstruct spatial communication, Ens et al.'s recent literature review showed that studies in this area neglect real-world scale complexities \cite{ens2019revisiting}. Although focusing on problems isolated to dyads helps researchers control their study environment, the intersecting complexities of scale produced by large-group collaboration are important for supporting real-world interactions. For example, in scaled collaborative events, multiple users with multiple roles might join from various devices or spaces. These spaces may scale across many different locations or a few locations with co-located users. Depending on the user's available space, these spaces might be small or large. Failing to explore these intersecting scale complexities poses a risk to the field's understanding of collaborative MR. 

As such, in this paper, we sought input from a class of professionals who are specialists in designing collaborative experiences for the physical world---\textbf{collaboration designers} (CDs). CDs are professionals with extensive tacit knowledge about how people collaborate and regularly create spaces to facilitate large-scale collaboration. Through a series of design explorations, we draw implications for extending their insights to collaborative experiences in  MR. By engaging them, we aim to surface new considerations for distributed group collaboration in MR. This approach takes inspiration from prior design work that recruited experts and/or professionals to validate the direction of future work on emerging technologies~\citep{reilly2010space, rajaram2023eliciting}. Our overarching research question is \textbf{how might collaboration designers re-imagine distributed mixed reality spaces for large-scale collaboration?}

To explore this question, we conducted a three-step expert workshop with ten CDs. The user study was specifically designed to (1) familiarise CDs with MR technology, (2) surface design principles that guide CDs' when they design collaboration spaces, and (3) elicit blended collaboration patterns to inform how to extend recent MR collaboration approaches to large groups of users. To do so, we (1) guided CDs through a technology probe experience, (2) performed a semi-structured interview, and (3) engaged CDs in a prototyping activity. We used qualitative affinity mapping to review the data and developed a series of collaboration space design principles and blended collaboration patterns (BCP) that can be built and tested as options to scale MR applications for distributed collaboration.

From our findings, CDs' space design choices can be themed into three key areas: keeping collaborators in a flow state; influencing the way collaborators feel, act or think; and designing the environment so it can respond dynamically to actions in the space. These informed the prototypes CDs created, which were synthesised into eight repeatable BCPs for scaled collaboration in distributed MR. 

These BCPs support four design implications for scaling MR collaboration. Firstly, MR creates new types of \textbf{blended facing formations (f-formations)}, which, unlike Kendon's original f-formations, \textit{blend the physical world with digital avatar f-formations}. This inclusion of the physical world, along with multi-user interactions across distributed environments, complicates how MR systems render user interactions.  Secondly, CDs consciously layer places on places to communicate non-verbally, but further work is required to understand how enabling blended places supports space alignment in distributed MR. Thirdly, interactions between large-scale, distributed groups require designers to think from vantage points other than the first person. Finally, MR needs a new design language that articulates how proxemics is supported across multiple blended physical spaces, with varying symmetries and alignment.\newline

\textbf{This work advances three major contributions:} 
\begin{itemize}
    \item A set of collaboration space design principles, based on professional practice, that researchers can use as prompts for future collaboration space studies.
    \item   Practice-informed blended collaboration patterns that can be used as a starting point for future research to scale MR systems from remote dyads to distributed collaboration spaces and many-to-many user interactions.
    \item Commentary on f-formation and place/space theories; how these ideas are used in MR to create blended f-formations and blended place-making.
\end{itemize}

The results of this work demonstrates the need for researchers to think beyond isolated scale problems and explore complex group work that requires trade-offs between spatial communication and visibility. By engaging CDs we contribute a series of opportunities, grounded in existing large-scale collaboration practices, for future evaluative work to expand distributed MR to real world collaborative events. 

\section{Related Work}

Mixed reality (MR) technologies allow distributed participants to communicate spatially. However, limited research examines how scale complicates interactions in collaborative MR \cite{ens2019revisiting}. Scale can be understood from three vantage points: multi-users and roles, mixed presence and space, or surface architecture \cite{ens2019revisiting, speicher2019mixed}. Unlike prior work, our research observes how these categories intersect because these challenges do not happen in isolation during real-world collaboration. 

To explore how MR spaces handle real-world collaboration, we engaged a group of collaboration designers (CDs). This specialty group of practitioners regularly design spaces and observes how participants turn these into places for collaboration. We took inspiration from prior work in expert participatory design \cite{rajaram2023eliciting, reilly2010space, zhou2023here} to help us observe how CDs apply their tacit knowledge to develop large-group interactions for MR spaces. Theories of proxemics and blended f-formations (facing formations) create a frame for examining these prototypes from a human-computer interaction (HCI) perspective. This sets the stage for problems that HCI researchers may wish to tackle in future evaluative research. 

\subsection{Large-scale distributed collaboration in mixed reality (MR)}
The ability of MR technologies to flexibly align real and digital objects across physical and virtual spaces \cite{milgram1994taxonomy, speicher2019mixed} makes them an ideal medium for distributed collaboration. Unlike large multi-display environments, such as media spaces \cite{buxton2009mediaspace, reilly2010space} or blended interaction systems \cite{o2011blended}, MR allows collaborators to digitally match misaligned spaces \cite{congdon2018mergingEnvironments, fink2022relocations, gronbaek2023partially, pejsa2016room2room, Ort2016holoportation} or warp avatars' poses and movements so they point or turn in the `correct direction' \cite{yoon2021full, wang2022avatarmotion, jo2015spacetimeAvatar}. These embodied interactions \cite{dourish2001action} 
open new opportunities for improving remote collaboration because users can intuitively communicate using interactions so far only available in person, like pointing or moving toward an object \cite{benford1998understanding, lawson2007language, buxton2009mediaspace, muller2017landmarks, pejsa2016room2room}. Kiyokawa et al. \cite{kiyokawa2002communication} found that a collaborative system with unmediated communication produced the least amount of miscommunication and that as non-verbal communication becomes more difficult, users tend to compensate with speech. This suggests the closer distributed MR gets to feeling co-located, the better people will perform in collaborative tasks. Further,  collaborative interactions can be grounded in the physical room, providing natural haptic feedback when the user sits at a desk or draws on a whiteboard \cite{gronbaek2023partially, o2011blended, reilly2010space}. Gr{\o}nb{\ae}k et al. \cite{gronbaek2023partially} showed that users reported a feeling of co-presence when the remote users' avatars were aligned with shared physical objects in MR. This ability to spatially communicate across dissimilar, distributed locations in a co-located way makes collaboration one of the `killer applications' \cite[p.g. 2]{ens2019revisiting} for MR. 

While distributed collaboration is a strong use case for MR, there is limited research in MR that investigates interactions at scale \cite{ens2019revisiting}. From the available literature, we focus on the intersection of three scale types: (1) the size of the collaborating group and member roles \cite{ens2019revisiting}, (2) the mixed presence of the group \citep{ens2019revisiting} and (3) the surface or space architecture blended within the MR system \cite{speicher2019mixed}. In their 2019 literature review of collaboration in MR, Ens et al. found that most studies ``did not match the complexity of real-world collaborative tasks'' \cite[p.g. 28]{ens2019revisiting}. This presents a challenge as situations, such as large-group workshops, conferences, or events often need to manage scale in intersecting ways. For example, large-group collaboration with many-to-many participation and various roles may cater for multi-modal ways of joining and might create larger physical and/or digital spaces to house participants. While there are studies that examine many-to-many interactions in MR \cite{yan2023coneSpeech, kasahara2016parallelEyes}, multi-modal and mixed presence collaboration \cite{Irlitti2023volumetric, schroder2023asymmetricDyads} or scaled spatial architecture \citep{speicher2019mixed}, these evaluative studies either manage complexity by making participants co-located \cite{kasahara2016parallelEyes, speicher2019mixed} or isolate a particular scale problem to evaluate \citep{schroder2023asymmetricDyads, Irlitti2023volumetric}. 

While a clear scope is important for scientific evaluation, our work deliberately observes the messy intersection of different scale challenges. In doing so, we identify ideas for future research that may have been overlooked by studying scale challenges in isolation. 

\subsection{Engaging collaboration designers to visualise scaled remote collaboration}
In their influential essay on ``re-placing-space", Harrison and Dourish emphasise the need for designers of digital spaces to understand \textit{place-making} \citep{harrison1996re}. While space and place have a complex relationship in HCI \cite{dourish2006re, krogh2017sensitizing}, involving \textbf{collaboration designers }(CDs) in the design process can help bridge the divide between space design and place-making. The idea of place-making is supported by the influential architect and scholar Bryan Lawson \cite{lawson2007language}, who explains that the way people act in space changes it. For example, a stadium without people is the \textit{same space} but a very \textit{different place} during a football match. In order to create purposeful and functional MR collaboration spaces, it is important to understand how participants transform these \textit{spaces} into \textit{collaboration places}. However, on revising the space-place problem ten years later, Dourish notes that ``space is just as much a social product as place" \citep[p.g. 301]{dourish2006re} because designers use their own understanding of the world to create the physical containers that become places for interaction. Krogh et al. \citep{krogh2017sensitizing} lean into this idea that ``it is time perhaps to re-space place" \citep[p.g. 306]{dourish2006re}, opting to focus on the way we intentionally design physical elements in MR to create optimal places for purposeful interactions. Inspired by this shift to hone in on space design in MR, we purposefully examine the creation of MR collaboration containers. Still, we aimed to retain an element of place-making by involving CDs, who specialise in how collaborators turn spaces into places for collaboration. 

In their practice, CDs pay special attention to how people interact with both digital and physical spaces in their work. Our research targeted CDs who use the MG Taylor methodology \cite{mgtaylor}, which can be described as ``...a method-of-methods—a uniquely modular approach that practitioners use to design and facilitate bespoke large-group collaborative design interventions"  \citep[p.g. 6]{brooks2019towards}. The MG Taylor method, based on years of practice and research since its first inception in 1979 \cite{mgtaylor, brooks2019towards}, focuses on the ``...integration of the physical environment, work processes and technology augmentation to facilitate human creativity and GroupGenius®" \cite{taylor2008}. Informed by the MG Taylor practices, Coullomb and Collingwood-Boots explain space design and facilitation as ``...the most obvious lever that contributes to the experience of the participants" \cite[p.g. 180]{coullomb2017collaboration}. CDs specialise in this intentional design of ``the metastructure of space, the visual component, the auditory ambience and the physical experience..." to create ``harmony and synchronicity" during large-scale collaborative events \cite[p.g. 186]{coullomb2017collaboration}. This approach to space design and place-making provides new practice-informed perspectives for how MR spaces can be designed to support distributed collaboration at scale. 

CDs design spaces to facilitate collaborative work, which moves dynamically between individual, group and whole of room situations. This allows participants to generate deep insights as individuals before sharing and synthesising their formulated ideas as a larger group. This dynamic view of collaboration aligns with Gutwin and Greenberg's \cite{gutwin1998designforgroups} exploration of \textit{mixed-focused} collaboration where ``people frequently move back and forth between individual tasks performed in relative isolation and shared work undertaken with other members of the group" \cite[p.g. 207]{gutwin1998designforgroups}. Our study builds on this holistic treatment of collaboration and observe how CDs design spaces that support the dynamic switching between individual, group and whole of room situations.

To engage these CD professionals, we took inspiration from several MR or collaboration space studies that involved expert user studies. For example, in Rajaram et al.'s \cite{rajaram2023eliciting} methodology, experts in augmented reality (AR) and security and privacy were recruited for an elicitation study to understand how designers can build security and privacy measures into AR tools. Their approach helped them to control assumptions made by AR experts that might sway the quality of their security and privacy research outcomes. Inspired by this, our research incorporated CD experts to control assumptions that HCI researchers might make about MR collaboration spaces. In terms of designing distributed spaces for collaboration, Reilly et al. \cite{reilly2010space} engaged experts in different design fields, including product design and architecture, to co-design a media space for remote collaboration. While their work showed promise, technological limitations challenged their ability to create blended physical and digital spaces. Our research revisits their idea of involving design experts, using advancements in MR to bridge their technical limitations and create blended collaboration environments. In a more recent study, Zhou et al. \cite{zhou2023here} recruited expert dancers to try a technology demonstration and co-design future applications for a smart mirror that assists improvisation. Similarly, we used an MR technology probe \cite{hutchinson2003techProbes, gaver2004driftTable} to inspire CDs and help them envision how they might use MR to create spaces for distributed collaboration. Involving experts in the design process helped Zhou et al. \cite{zhou2023here} to provide an experienced evaluation of the ideas dancers suggested in their user study. This type of expert involvement is particularly helpful in reviewing research in skill areas that require a high level of tacit knowledge, such as dance or in our case, group collaboration.  

\subsection{Proxemics in mixed reality}
Proxemics is often used as a framework for understanding and designing computer-supported group interactions \cite{gronbaek2023partially, Irlitti2023volumetric, o2011blended, mentis2012interaction}. The architect Edward T. Hall coined the term proxemics while exploring how people use space as an extension of culture \citep{hall1966hidden}.  For Hall, the dynamics of people and objects in space play an important role in non-verbal communication  \citep{hall1966hidden}. For example, social intimacy can be non-verbally communicated using the distance between two people \cite{ hall1966hidden}. HCI research applies this understanding of spatial interactions between people and objects to support group collaboration \citep{mentis2012interaction, gronbaek2023partially, dourish2001action, greenberg2011proxemic}. Initial research used proxemics to automate computer responses according to intimate, social, and public tasks \cite{greenberg2011proxemic}. Since then,  collaborative computing increasingly borrows from proxemics to design systems that help actors communicate information and coordinate collaborative activities through their spatial interactions \cite{gronbaek2017proxemic, Irlitti2023volumetric, mentis2012interaction, o2011blended}. 

\subsubsection{Blended F-formations}
\begin{figure}[ht]
\centering
\includegraphics[scale=0.3]{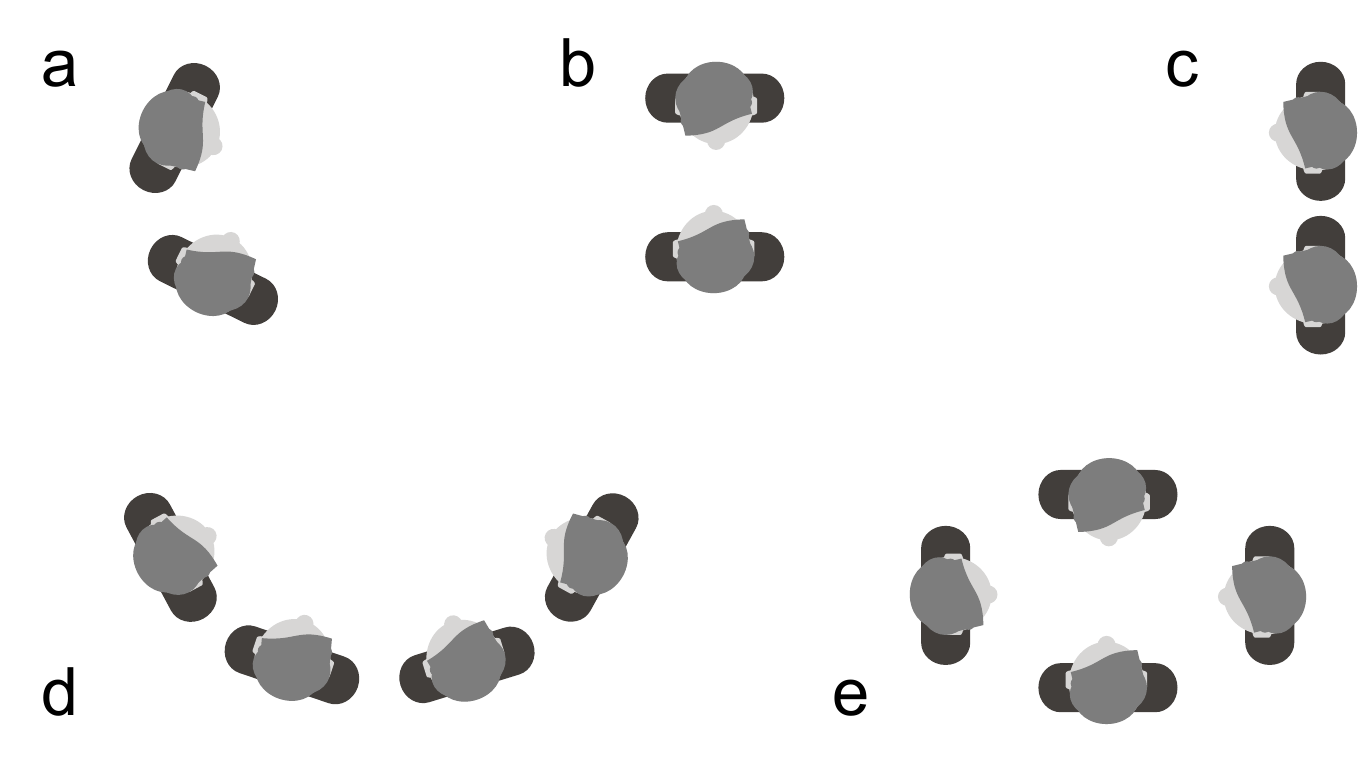}
\caption{\textbf{Adam Kendon's F-formations: \newline a. L-arrangement; b. face-to-face; c. side-by-side; d. semi-circular; e. rectangular \citep{marshall2011f-formation, kendon2010spacing}}}
\label{fig:Kendon_f-formations}
\end{figure}

Adam Kendon's observational theory of facing formations (f-formation) illustrates how people organise themselves in relation to others and artefacts in space \citep{kendon1990conducting}. An f-formation is created when participants in a collaborative environment ``...organise themselves so that a shared transactional space is established and maintained" \cite[pg.5]{kendon2010spacing}. These formations typically have 2-5 participants \citep{marquardt2012cross} and can take a range of shapes, depending on the number of participants and required task \citep{marshall2011f-formation} \textbf{(figure \ref{fig:Kendon_f-formations})}. For example, a square shape might be formed when four or more people face each other to share in a conversation. The inner circle of the f-formation (o-space) is the group's central focus, where activities occur. The spaces between participants (p-space) ``determine group membership" \citep{marquardt2012cross}. Finally, the surrounding space (r-space) wraps around the group, delineating them from the surrounding environment. The group monitors this space and will notice when others enter this space, attempting to join. 

Studying f-formations in a variety of contexts has enabled researchers to surface new mechanisms for interaction \citep{brudy2018eagleview, marshall2011f-formation, mentis2012interaction, paay2013f, tong2016s}. Although we do not directly observe f-formations, our study examines how participants design MR spaces to facilitate what we call \textit{blended f-formations}. These blended f-formations differ slightly from Kendon's original f-formations in that they involve Interaction Proxemics \cite{o2011blended, mentis2012interaction} with objects in the environment. For example, in ClearBoard \cite{Ishii1992ClearBoard}, participants form a face-to-face f-formation (Figure \ref{fig:Kendon_f-formations}-b), yet an interactive surface acts as a participatory object blended between them in the f-formation. Understanding blended f-formations helps to reveal moments in the CDs' MR prototypes where objects are blended with group activities to support seamless collaboration.

\subsubsection{Blended object classifications }
The way objects move in space creates the scaffolding for blended f-formations. Inspired by Hall and Marquardt et al., we classify objects as fixed, semi-fixed, or mobile when discussing their spatial dynamics \citep{marquardt2012cross, hall1966hidden}. \textit{Fixed} objects are parts of the space that cannot be or are rarely changed, for example, a wall-mounted whiteboard. This contrasts to \textit{mobile} objects, which frequently change spatial configuration, like a person's tablet. \textit{Semi-fixed }objects sit between these two ends of the spectrum and describe objects that are re-configurable but not completely fixed or mobile. An example of a semi-fixed object might be a chair or whiteboard on wheels. Prior work in MR and blended interaction spaces has explored blending fixed and semi-fixed objects in collaborative activities \citep{gronbaek2020proxemics, o2011blended, Ishii1992ClearBoard, marquardt2012cross}. However, animating transitions between physically misaligned fixed objects may be challenging when we scale remote collaboration. We are interested in how CDs ideally blend fixed, semi-fixed and mobile objects with f-formations and identify problems in these interactions at scale.

\section{Study Design}
Spatial communication systems are complex and influenced by culture, both from the designer's perspective and the subjective end user's interpretation \citep{dourish2001action}. This work deliberately draws on collaboration designers' foundational understanding of how people interact in physical and digital collaborative spaces, which is an important lever in their MG Taylor informed practice of work \cite{coullomb2017collaboration}. 

\subsection{Method}
We recruited ten expert participants for parts 1-3 of our study (n = 10, median experience = 7 years, average experience = 6.9 years, min experience = 2 years, max experience = 12 years) and six for part 4 of the study (n = 6, median experience = 9 years, average  experience = 8 years, min experience = 2.5 years, max experience = 12 years). Our inclusion criteria specified participants must be currently working in a role that involves designing, facilitating, creating, or enabling a space (digital or physical) that helps four or more people work collaboratively. We used a snowball sampling technique to recruit participants from this close network of professionals. Each participant read a plain language statement and signed a consent form before taking part in the study. 

At the start of each one-on-one in-person session, the participant answered a survey about their skill background and experience using MR tools. A summary of each participant's core skills, years of collaboration experience and familiarity with MR tools can be found in the appendix. 

\begin{figure}
    \centering
    \includegraphics[width=1\linewidth]{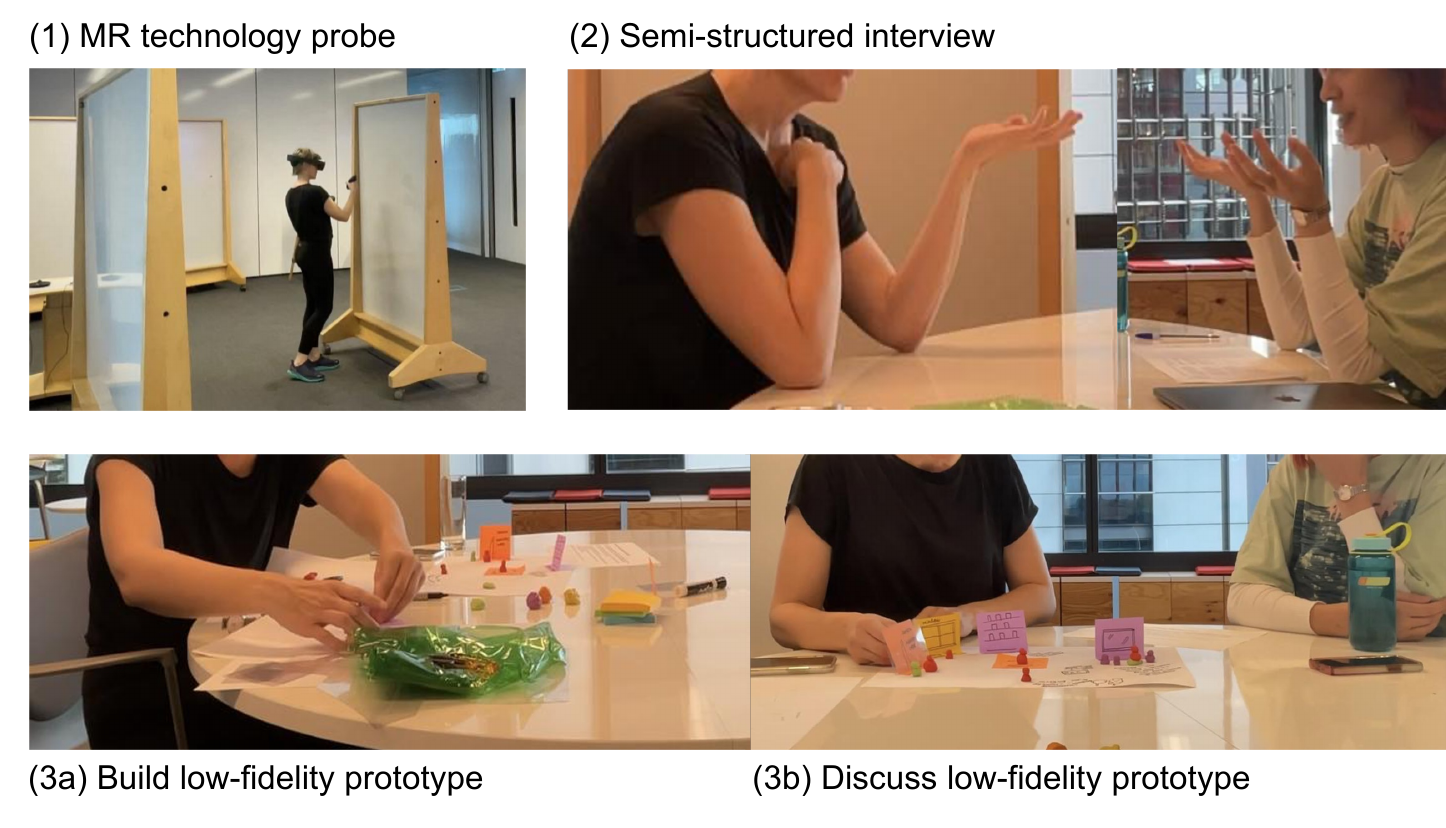}
    \caption{\textbf{The expert workshop involved (1) a MR technology probe, (2) a semi-structured interview to understand the way CDs design physical and digital collaborative spaces, (3a) a low-fidelity speculative prototyping activity and (3b) a discussion about the prototype they made.}}
    \label{fig:Method2}
\end{figure}

To capture and qualitatively analyse our CDs' space design choices, we carried out a three-part expert workshop \textbf{(Figure \ref{fig:Method2})} and follow-up validation session. The overall design brief of the workshop was for CDs to ``brainstorm ways [they] would adapt MR technology to design and facilitate a virtual event. By this, we mean an event where the participants are in a physical space, wearing a headset to collaborate with each other". Participants were free to use a blend of AR and VR capabilities, which follows Milgram and Kishino's Reality-Virtuality continuum definition of MR \cite{milgram1994taxonomy}. Participants were also free to create solutions for distributed scenarios, where no participants are co-located, or hybrid scenarios, where some participants may be co-located. The following sections detail each part of the study and the intent behind these activities. 

\subsubsection{Part 1: MR technology probe}
Though CDs are highly knowledgeable in designing collaborative work, they may have limited experience using MR. This could potentially limit their ability to brainstorm MR-based solutions. Zhou et al. \citep{zhou2023here} overcame a similar issue by creating a technology probe-like demonstrator for their recruits to experience MR mirror technology. 

Inspired by this approach, we addressed this challenge by designing an opportunity for CDs to experience collaborative MR. As a technology probe, we used a simplified version of Grønbæk et al.'s \textit{Blended Whiteboard}~\cite{gronbaek2024bw}. The application runs on the Meta Quest Pro headset and allows users to experience passthrough MR in full colour, i.e. when wearing the headset, participants see the physical room around them in colour, with a digital overlay for blended objects. Looking down at their hands, they see the graphic controller options and their avatar's cartoon sleeves over their arms.

\begin{figure}
    \centering
    \includegraphics[width=1\linewidth]{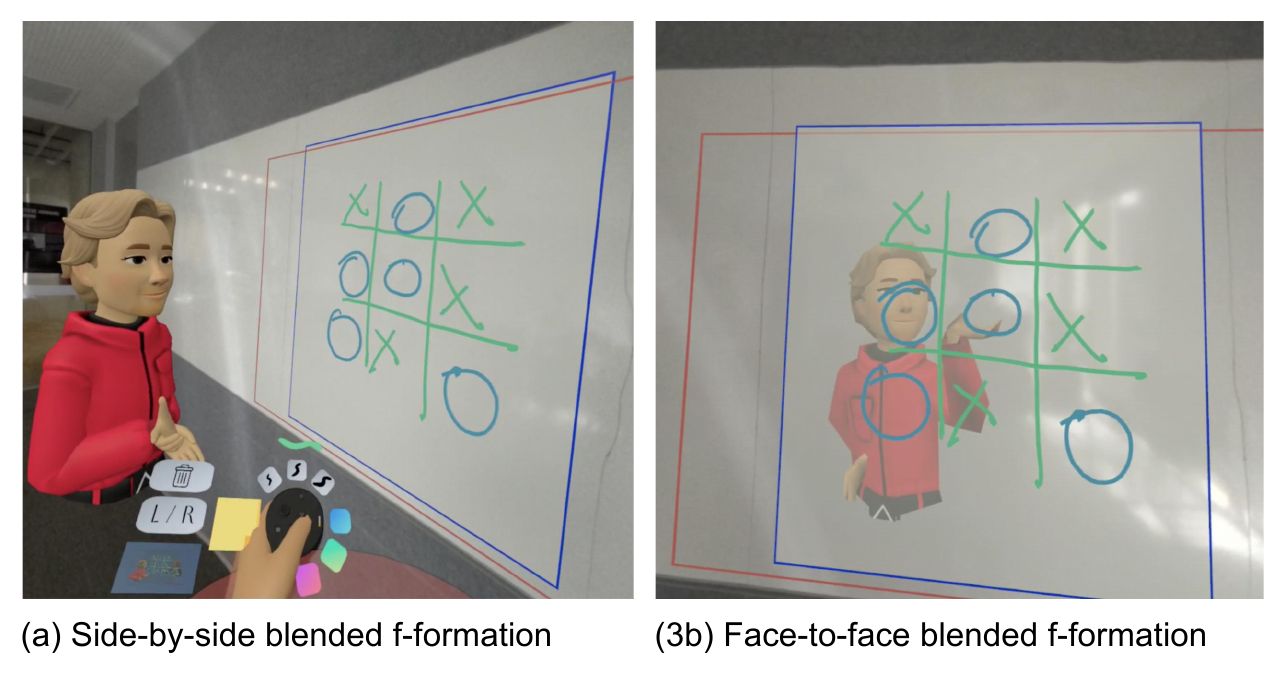}
    \caption{\textbf{Example of a game of tic-tac-toe played by participants during the technology probe stage. It demonstrates the two different blended f-formations: (a) side-by-side where the blended whiteboard is in front of the partners and (b) face-to-face where the blended whiteboard is between the partners.}}
    \label{fig:Method3}
\end{figure}

We used this technology probe to showcase two key MR spatial configuration features that are difficult to explain unless experienced. Firstly, participants can define collaborative objects in the room, such as a desk or whiteboard, as an anchor for aligning the distributed room configurations. This approach supports common collaborative f-formation configurations, face-to-face and side-by-side \cite{kendon1990conducting, Ishii1992ClearBoard, He2020CollaboVR}. In face-to-face mode, the partner's avatar is displayed inside the whiteboard surface, similar to \textit{ClearBoard's} \cite{Ishii1992ClearBoard} blended face-to-face f-formation. In side-by-side mode, the partner's avatar is positioned outside the whiteboard, floating beside the user \textbf{(Figure \ref{fig:Method3})}. We included these features in the technology probe so that participants could have a more complete experiential understanding of how these difficult-to-explain spatial configuration features work. The intent was to help CDs feel more comfortable designing with these interactions, having experienced them. 

Before trying the technology probe, the participant listened to a brief explanation of the system and watched a video of it being used. This helped to familiarise them with the controls before they tried on the headset. The study space was set up so the interviewer and participant faced opposite directions when standing in front of a whiteboard. A dividing wall was placed between the participant's space and the interviewer's so that they could not see the other person's physical body during the technology probe. 

We used a game of tic-tac-toe to demonstrate the different blended f-formations in the technology probe \textbf{(Figure \ref{fig:Method3})}. We selected this familiar game because it involves drawing, turn-taking, and interpreting shapes, which are common collaborative interactions. 
\subsubsection{Part 2: semi-structured interview}
After the technology probe, we conducted a semi-structured interview with each participant to understand how they approach the design of spaces for collaboration in their professional practice. The open-ended questions focused on participants' collaboration space design and decision-making processes. Probing questions were asked when interesting spatial ideas were discussed. This method was chosen in order to explore experiences that inform their space designs \citep{chennamaneni2011integrated}. 

\subsubsection{Part 3: low-fidelity prototype}
To capture CDs' design ideas for MR remote collaboration spaces, we gave participants an opportunity to prototype their ideas. Prototyping is a common method used to communicate ideas in HCI and can involve a variety of materials \citep{koskinen2011design}. We specifically chose tools that would prompt participants to think about how collaborators would position themselves in relation to other avatars or objects and how these objects might overlap and blend. These tools included modelling clay, sticky notes, felt pen, perspex, cellophane, paper, whiteboard markers and scissors. All participants had prior experience with low-fidelity prototyping.

\begin{figure}
    \centering
    \includegraphics[width=1\linewidth]{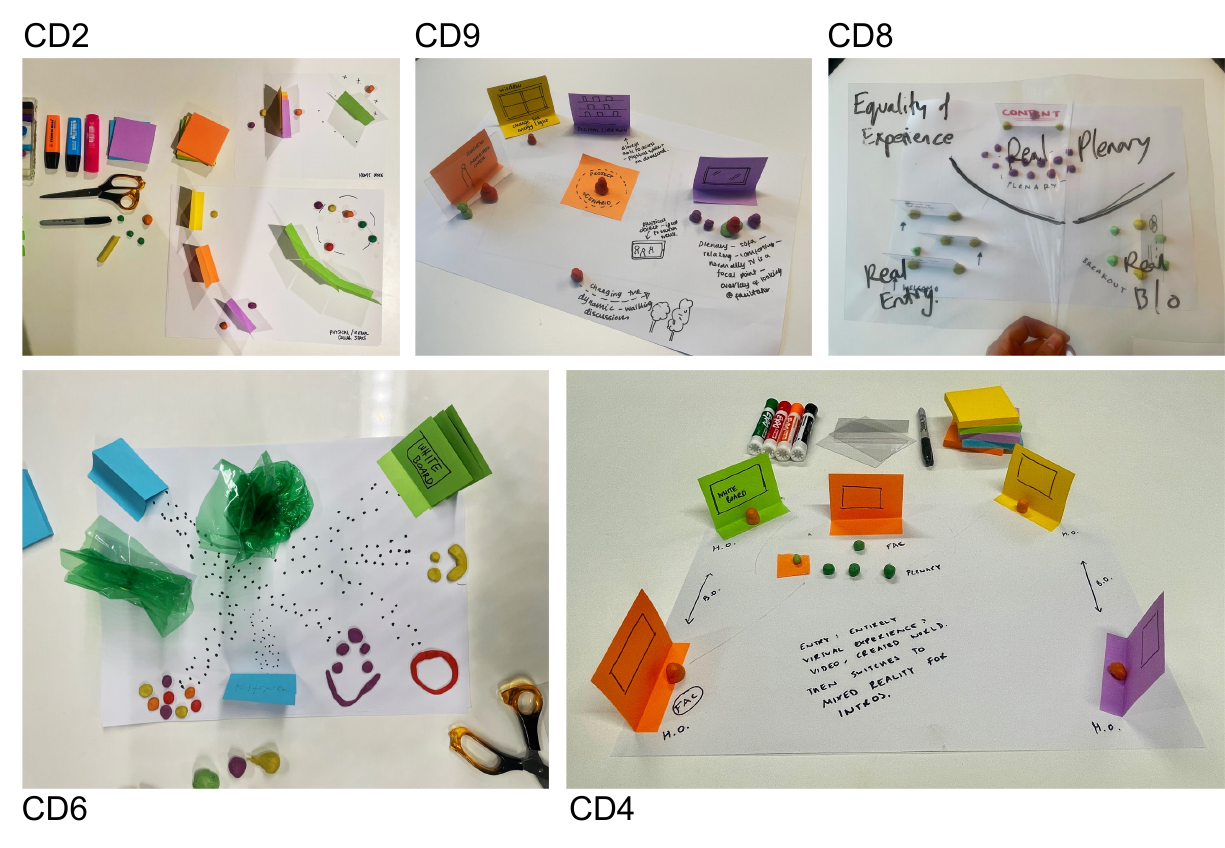}
    \caption{\textbf{A selection of speculative low-fidelity prototypes created by collaboration designers 2, 9, 8, 6 and 4 (as pictured left to right).}}
    \label{fig:Prototypes}
\end{figure}

The participant had approximately ten minutes to construct their prototype using the tools provided. The prompting questions for this activity can be found in the appendix. Once they were happy with their prototype, the interviewer rejoined them and asked the participant to explain their design, probing further if they talked about an interesting spatial idea. \textbf{Figure \ref{fig:Prototypes}} shows an example of prototypes.

Although common in participatory design methods, we did not use a think-aloud protocol. This was because we wanted to give participants time to think deeply about their designs without adding a verbal cognitive load \citep{jaaskelainen2010think}. Post prototyping, their designs were discussed in-depth, in an unstructured way, to fully explore the decisions they made about their collaborative MR space.
\subsubsection{Part 4: Validation}
After synthesising the first round of results, several weeks post the first user session, we validated the collaboration space design principles with 6 CDs (n = 6, median experience = 9 years, average experience = 8.1 years, min experience = 2.5 years, max experience = 12 years). 4/6 of the CDs were involved in the prototyping activity and 2/6 were new CD recruits, as some of the original CDs were unable to participate in  the validation session due to work commitments. The validation comprised of six 30-minute one-on-one semi-structured interviews, where the interviewer gathered feedback from CDs on whether the collaboration space design principles reflected how they think about space design in their work. All participants signed a consent form, and transcriptions were made of the voice-recorded interviews using a research tool called \textit{DoveTail}\footnote{https://dovetail.com}. After the third participant, the design principles were iterated according to the transcripts, and the last three participants provided feedback on the second iteration of the design principles. This last round of feedback was also incorporated and the final principles are presented in the results section. 

\subsection{Data Collection and Analysis}
The qualitative study collected both audio and visual data. The audio data was transcribed using \textit{DoveTail}. Affinity mapping, which incorporated both deductive and inductive coding, was used to show patterns in the interview data. Once all the data was coded, the tags were reviewed, merged, or changed where applicable. These were then separated into technology probe, interview, and prototype parts and grouped into themes using the automated affinity mapping feature in \textit{DoveTail}. Photos of the participants' low-fidelity prototypes were visually analysed to identify patterns in their designs. These themes and patterns were then reviewed and refined into insights.

This approach surfaced two sets of insights; collaboration space design principles and blended collaboration patterns (BCP). The collaboration space design principles show the guiding principles CDs use to develop digital and physical spaces for collaboration. These design principles were synthesised using affinity mapping of the interview transcripts. Once synthesised, the design principles were shared with CDs during a semi-structured interview and further refined based on their feedback. The blended collaboration patterns (BCP) are a set of design insights that highlight patterns in the way CDs designed MR collaborative spaces for distributed collaboration at scale. These patterns were first developed using affinity mapping of the interview transcripts, then cross-examined with patterns identified from the images taken of the participants' prototypes. Diagrams were created to visually communicate these themes so that readers could easily understand them and potentially build the key ideas CDs created as high-fidelity prototypes.

\section{Results}

To answer the overarching research question, we analysed the results with two sub-questions in mind:  
\begin{itemize}
    \item \textbf{RQ1: What principles do collaboration designers (CDs) follow when they design collaborative spaces? 
    \item  \textbf{RQ2: }What blended collaboration patterns do CDs identify when prototyping spaces enabled by collaborative MR? How are they informed by their practice?}
\end{itemize}

We present our findings in two steps. Firstly, in response to \textbf{RQ1,} we present the synthesised design principles to capture the strategies CDs use to create physical and digital collaboration spaces. We then use these as a reference for the second set of insights that answer \textbf{RQ2}. These insights take the form of blended collaboration patterns (BCP) that were thematically synthesised by observing patterns across the different prototypes. \textbf{Figure \ref{fig:DesignModules}} shows a visual representation of all eight blended collaboration patterns aligned to the design principles. 

These synthesised collaboration space design principles and BCP can be used by designers and researchers as a guide for future work. For example, designers might choose to print out the design principles and have these at hand when designing collaborative spaces. Alternately, researchers might use the BCPs to inform future empirical research that tests scaled collaboration in distributed MR, based on the BCP designs.

\subsection{Collaboration space design principles}
\begin{figure*}[ht]
\includegraphics[scale=0.6]{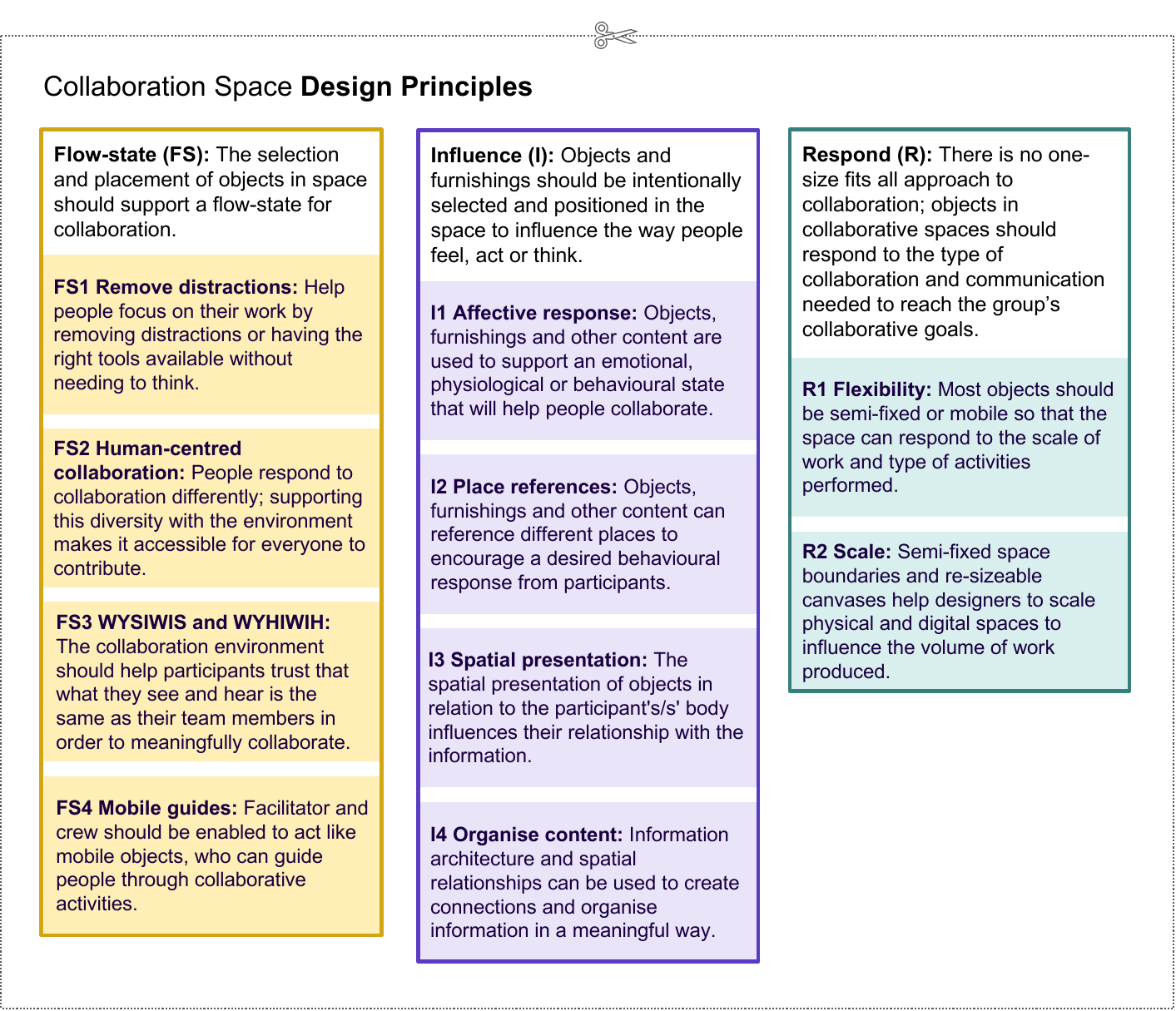}
\caption{\textbf{Collaboration space design principles synthesised by interviewing CDs about how they design physical and digital collaborative spaces. These can be categorised into three higher-order themes: \textsc{flow-state}, \textsc{influence}, and \textsc{respond}, with lower-order themes sitting below these.}}
\label{fig:DesignPrinciples}
\end{figure*}

The analysis surfaced a number of different ideas that we shaped into \textbf{design principles} in response to \textbf{RQ1}. We grouped them into three pillars: Maintain a collaborative \textsc{flow-state}, \textsc{influence} the way people feel, act or think and \textsc{respond} to their needs \textbf{(Figure \ref{fig:DesignPrinciples})}. When we validated the resulting design principles with the CDs, all participants agreed that these pillars were relevant, and no additional pillars were suggested. 

\subsubsection{Maintain a Collaborative Flow State (FS)}
CDs use different strategies that help them keep collaborators in a \textsc{flow state}. This is so their participants can \textbf{focus their energy} on the collaborative task. These strategies include \textsc{remove distractions} (FS1), where CDs intentionally remove items or processes that might distract from the collaborative task at hand, like positioning whiteboard markers in an obvious place so participants don't need to look for them; \textsc{human-centred collaboration} (FS2), designing the collaborative space so people feel it is safe and easy to contribute; what you see is what I see (\textsc{WYSIWIS}) and what you hear is what I hear (\textsc{WYHIWIH}) (FS3), ensuring participants have a shared view of how the group's ideas progress and acting as \textsc{mobile guides} (FS4) to help participants along their collaborative journey. 

\subsubsection{Influence Collaborators (I)}
CDs intentionally select and place items in collaborative spaces to \textsc{influence} how collaborators \textbf{feel, act or think}. The data showed that most CDs (9/10) mentioned how they design spaces to elicit conscious or subconscious responses from their collaborators. 

\begin{quote}
    \textbf{CD6:} \textit{``...we can \textbf{change the environment}, which \textbf{changes the way people think} and feel - well, changes the way they feel, which I think influences the way they think.''}
\end{quote}

These mechanisms can be categorised into four areas: \textsc{affective response} (I1), where CDs use the interplay of fixed and semi-fixed objects to elicit an emotional, physiological or behavioural response; \textsc{place references} (I2), the use of objects from specific places to communicate non-verbally (e.g. a playschool set used to encourage curiosity); \textsc{spatial presentation} (I3), placing information objects so that they hold extra meaning in the space and \textsc{organisation of content} (I4), using the spatial placement of objects to communicate information hierarchy. 

\subsubsection{Respond to Collaborator Needs (R)}
The data showed it is important for objects in collaborative spaces to \textsc{respond} to \textbf{collaborator needs}. CDs explained the responsiveness of the space is key to support solving a diverse range of collaborative problems. 

\begin{quote}
\textbf{CD10:} \textit{``...[the space] should be, be able to be \textbf{shifted} any number of ways depending on the really particular\textbf{ needs of the people} who are coming together and the problem that they're trying to solve.''}
\end{quote}

This responsive quality is enabled by the \textsc{flexibility} (R1)  and \textsc{scalability} (R2) of collaborative spaces, both with semi-fixed objects such as physically movable whiteboards and digital canvases. 

\subsubsection{Collaboration Space Design Principles Validation}
When we shared the resulting design principles back to the CDs, there were very few suggestions made to update the framework, and on the whole, the response was positive. As one participant noted, ``I would print this out and use it".

When asked how likely they would recommend using this framework to design collaborative spaces on a scale of one to ten, one being least likely and ten most likely, the median score was 8.5/10 and the mean 9/10. The only small changes were to include ideas such as accessibility, furnishings (e.g. wood or natural light) and sound. This suggests that the design principles synthesised from the semi-structured interviews accurately represent what CDs use to inform virtual and physical collaborative space design.

\subsection{Remote Collaboration Space Blended Collaboration Patterns (BCP)}
\begin{figure*}[ht]
\includegraphics[scale=0.5]{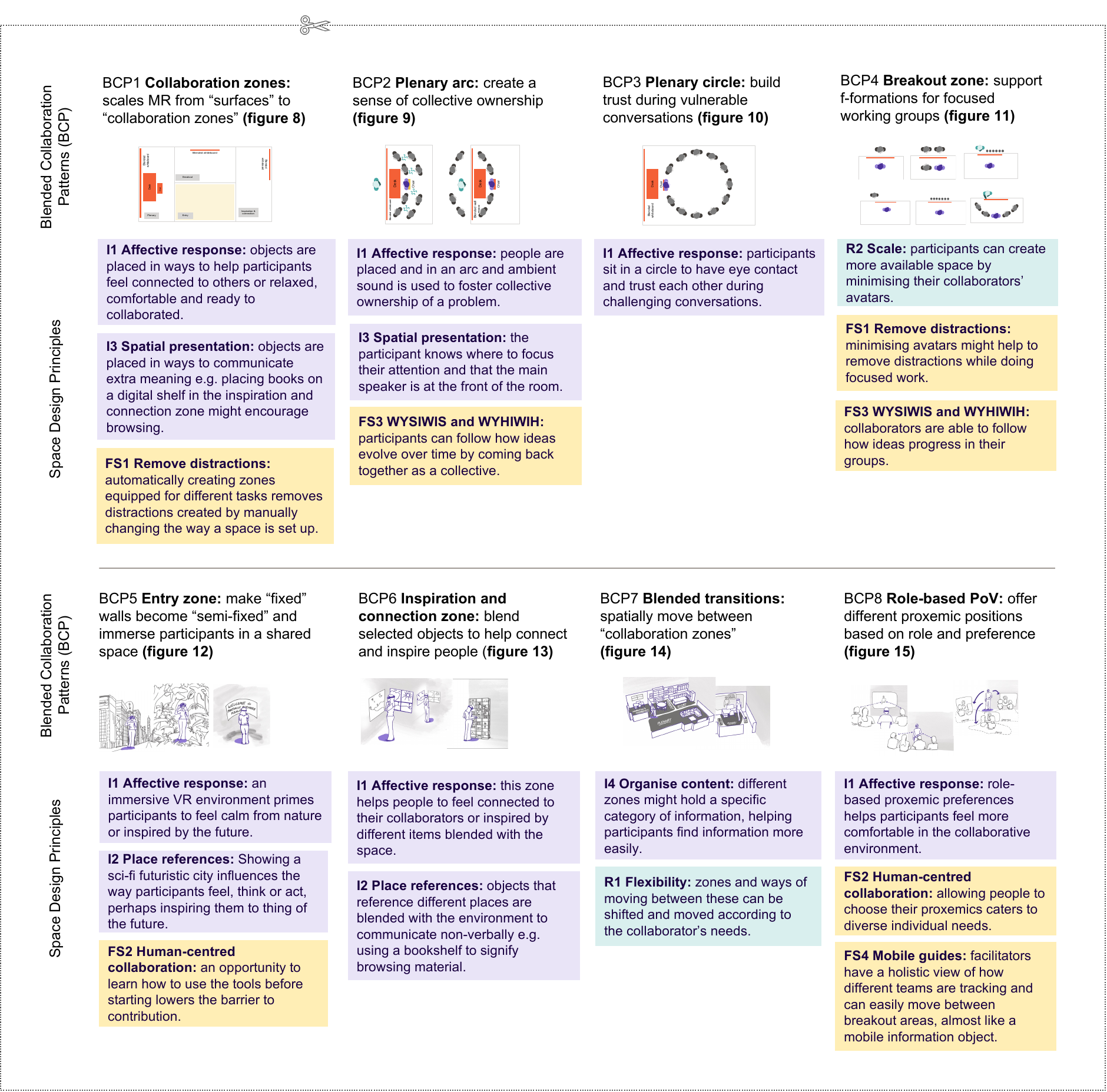}
\caption{\textbf{All eight blended collaboration patterns (BCP) thematically synthesised from the CDs' prototypes and mapped to the collaboration space design principles. BCP 1-6 show how CDs created different zones that facilitate, individual, group and whole of room activities. BCP 7 and 8 illustrates how participants might dynamically manage or move between these areas.} Each BCP diagram is shown in more detail later in the results.}
\label{fig:DesignModules}
\end{figure*}

In response to \textbf{RQ2}, we synthesised patterns observed from the prototypes into eight MR space \textsc{blended collaboration patterns} (BCP). These BCP resulted from CDs' tendency to create areas in their low-fidelity prototypes that catered to specific activities. These activities included variations of individual, multi-group and whole of room participation, which reflects their approach to collaboration as a holistic, dynamic process. The BCPs distill some of the novel and interesting approaches across all of the low-fidelity prototypes that have significance for scaling distributed MR. \rev{These provide practice-informed design options for researchers to use as a starting point to develop future empirical studies}. BCP 1-6 focus on the different areas or \textsc{zones} that CDs created, while BCP 7 and 8 present common ideas CDs had for managing or transitioning between these \textsc{zones}.  \textbf{Figure \ref{fig:DesignModules}} gives a visual overview of the BCP aligned to the design principle categories.

\subsubsection{BCP1: \textsc{Collaboration zones} scale MR surfaces to MR collaboration spaces}
\begin{figure}
    \centering
    \includegraphics[width=1\linewidth]{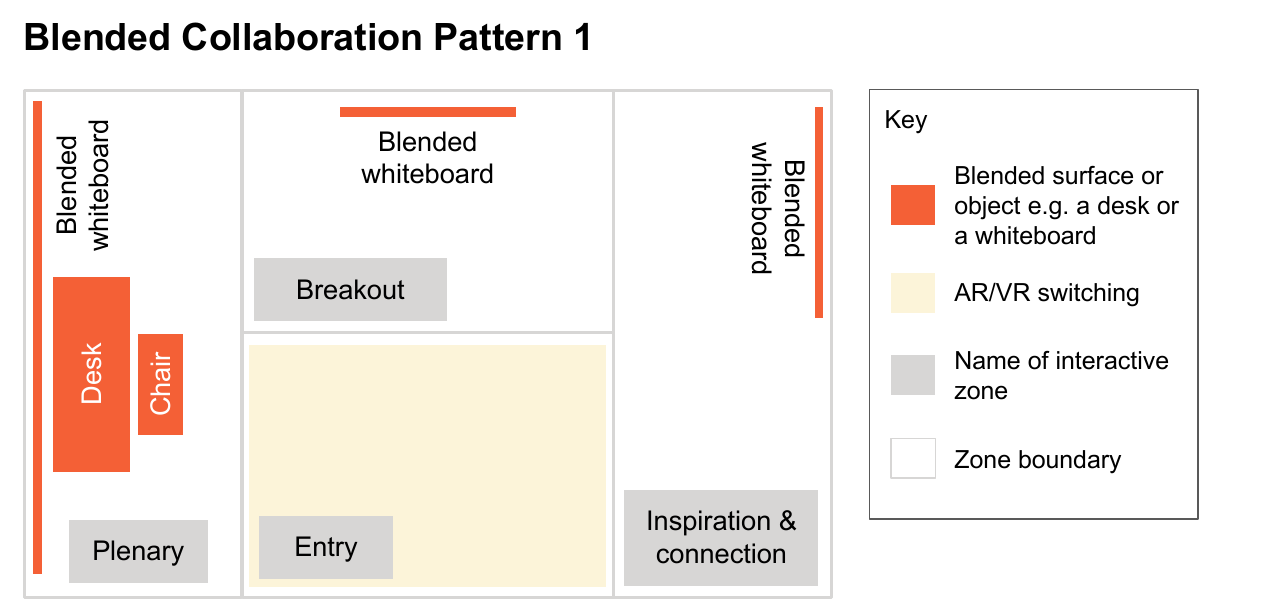}
    \caption{\textbf{BCP1) A birds-eye view of how a participant’s home office might be used to create different zones for collaboration. Each zone is set up for different collaborative purposes. For example, a breakout is used for smaller teamwork and the plenary is for sitting and collectively listening to content. The participant might transition between these zones by physically stepping into these different geo-fenced areas. In the entry area, users can switch between different AR and VR modes to create partially or fully immersive environments that prime them for collaboration. The particular spatial layout of these \textsc{zones} is just an example, but the role of each zone was consistently brought up by CDs.}}
    \label{fig:Zones}
\end{figure}

Throughout the user study, a pattern for \textsc{collaboration zones} emerged \textbf{(Figure \ref{fig:Zones})}. CDs naturally created these zones as areas that catered to different types of activities. These zones included:
\begin{itemize}
    \item    a \textsc{plenary zone} for listening to and sharing information as a collective, 
    \item      a \textsc{breakout zone} for working on tasks as a smaller group,
    \item an \textsc{entry zone} for preparing participants to collaborate 
    \item and an \textsc{inspiration and connection zone} to help participants think differently and get to know each other.
\end{itemize}

These zones each served different purposes that the results will explore further. These zones had different objects inside them to assist with the \textbf{particular tasks carried out in each zone}, following the design principles I3 and FS1.
\subsubsection{BCP2: \textsc{Plenary zone}: \textsc{plenary arc} create a sense of collective ownership}
Throughout the user study, CDs referenced using a \textsc{plenary arc} as a place for all members of the collaborative session to hear or work on information all together. A plenary can be described as the meeting of all participants in a working session. For CDs, this meeting of the collective is supported by the space, and all CDs talked about \textsc{plenary} as a dedicated meeting area. The main reason to include a \textsc{plenary arc} was to create a sense of \textbf{collective ownership} and \textbf{shared memory} since participants are all in the same area, next to each other, hearing the same thing. 

\begin{quote}
\textbf{CD4:} \textit{``...the advantage of plenary is you have a \textbf{collective experience}, and you have these really amazing moments and events where there's sort of like a breakthrough.''}
\end{quote}

\begin{figure}
    \centering
    \includegraphics[width=1\linewidth]{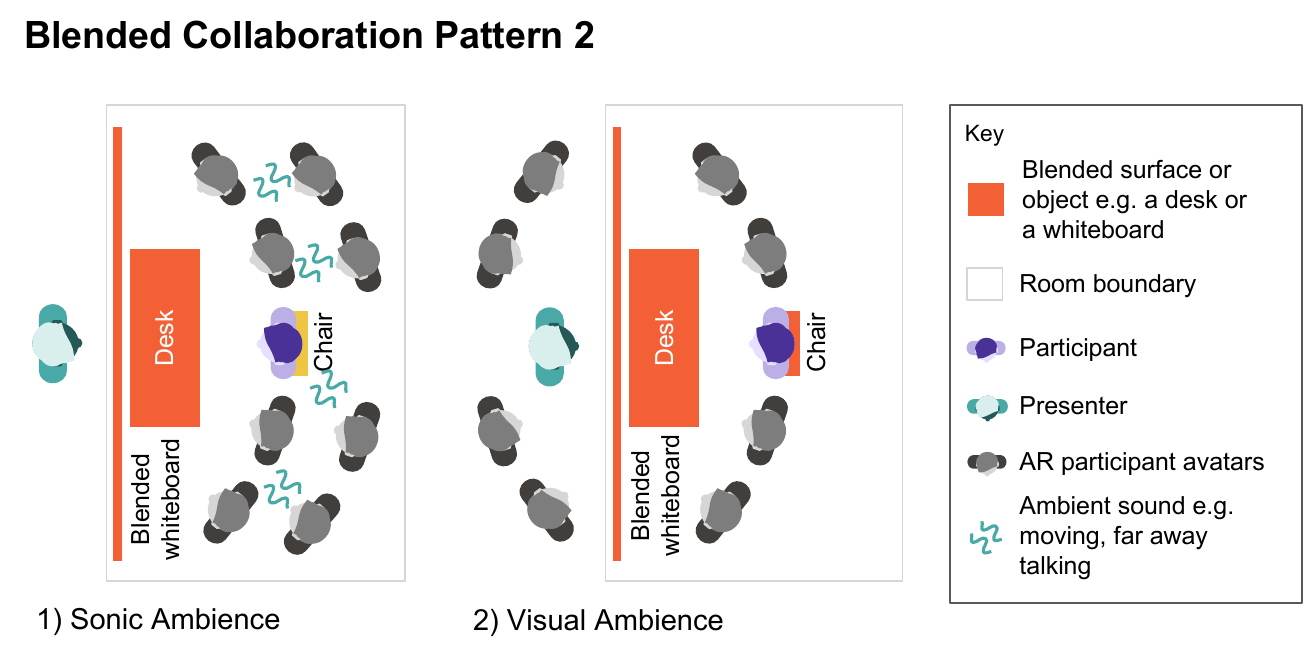}
    \caption{\textbf{BCP2) Birds-eye view of a \textsc{plenary arc}, as commonly suggested in the CDs’ prototypes. (1) Sonic ambience shows the presenter behind the participant’s blended whiteboard; they can ambiently hear other participants around them. (2) Visual ambience is similar, but this time shows glimpses of other participants inside their defined wall space.}}
    \label{fig:PlenaryArc}
\end{figure}

In this plenary area, CDs often described an arc seating arrangement \textbf{(Figure \ref{fig:PlenaryArc})}. CD1 likened this physical arc to the digital Microsoft Teams \textit{together mode}.

Predominately, CDs used the participant's chair as an anchor to create the arc \textbf{(Figure \ref{fig:PlenaryArc})}; the different avatars being automated to sit in this arc shape, facing a wall, maybe behind their desk, where the wall would act as a blended screen. The facilitator was often placed inside this wall, using the face-to-face blended f-formation. Interestingly, two separate CDs suggested including a \textbf{soundscape} to enhance the feeling of \textbf{`being together'}. 

\begin{quote}
\textbf{CD9:} \textit{``I think \textbf{soundscape} is really important, even like the \textbf{ambient noise}, to feel like you really are in a group of a hundred people. Being able to almost turn that on and off''}
\end{quote}

Similarly, CD3 suggested using the face-to-face blended f-formation to give participants \textbf{glimpses} of their fellow collaborators \textbf{(Figure \ref{fig:PlenaryArc})}, perhaps to make up for the lack of peripheral vision in the MR headsets. 
\subsubsection{BCP3: \textsc{Plenary zone}: \textsc{plenary circle} builds trust for collaborators to have vulnerable conversations}
\begin{figure}
    \centering
    \includegraphics[width=1\linewidth]{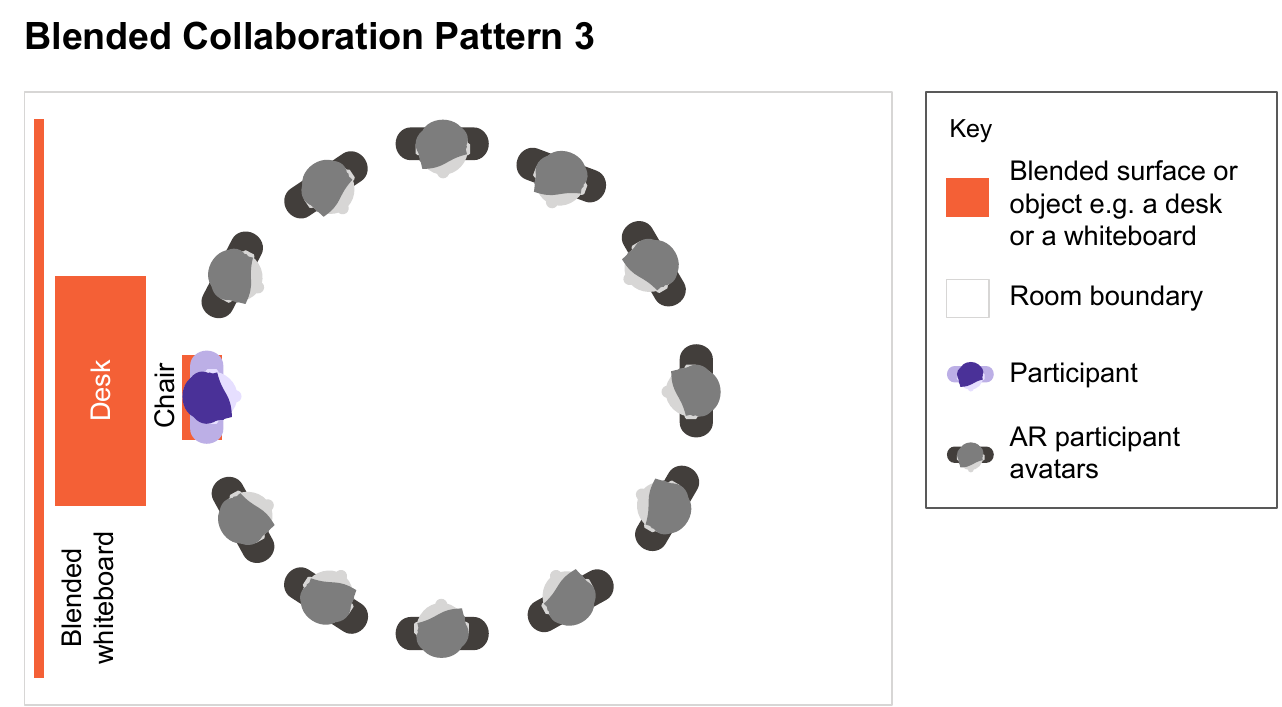}
    \caption{\textbf{BCP3) Birds-eye view of a \textsc{plenary circle}, as commonly suggested in the CDs’ prototypes. In this diagram, every participant would be seated in their chair at home and view their collaborators' avatars floating in a circle.}}
    \label{fig:PlenaryCircle}
\end{figure}

Similar to the \textsc{plenary arc}, CDs suggested including functionality in collaborative MR that sits participants in a circle. The main reason for this was to enable \textbf{eye contact} with other collaborators, which \textbf{builds trust} during vulnerable conversations \textbf{(Figure \ref{fig:PlenaryCircle})}.

\begin{quote}
\textbf{CD2} \textit{``...in a circle [...] it's a bit more of a, a time where we need to have a, you know,\textbf{ deeper honest conversation} around where we're heading, what we, what we're doing, what's getting in our way...''}
\end{quote}

Interestingly, CDs avoided placing objects inside the circle, as this would create a kind of \textbf{barrier} that breaks the feeling of being connected with other participants. 
\subsubsection{BCP4: \textsc{Breakout zone}: blended f-formations change according to the number of participants and level of interaction with diagrams shared on the whiteboard surface.}
Breakout areas help divide a large group of participants into \textbf{smaller teams} to do \textbf{focused work}. A smaller number of team members means each voice is more likely to be heard. 

\begin{figure}
    \centering
    \includegraphics[width=1\linewidth]{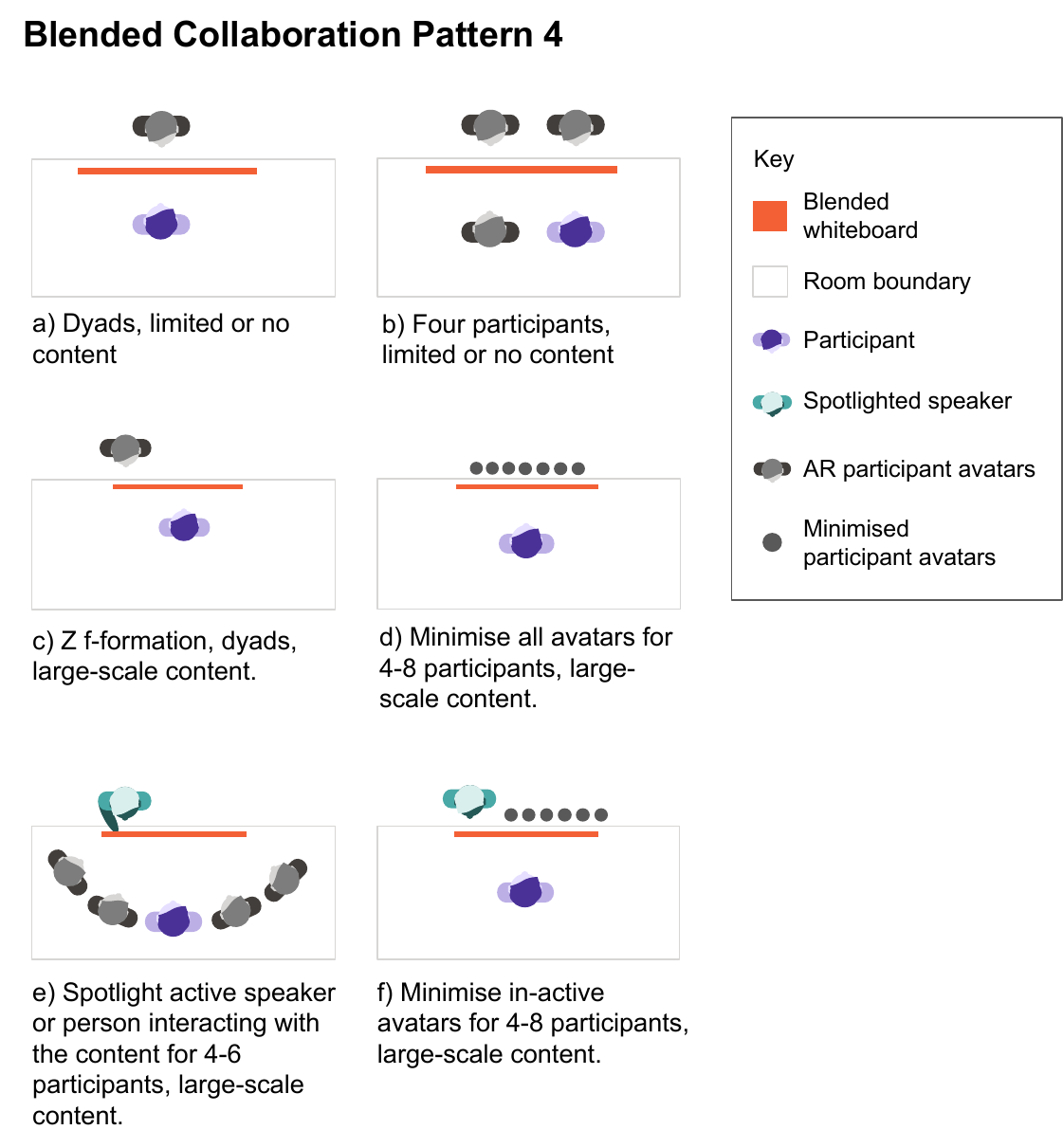}
    \caption{\textbf{BCP4) CDs suggested different ways to manage the number of participants in breakouts: (a - b) blended face-to-face formations often suggested for smaller breakouts, (c) Z-formation, which helps participants collaborate when there is a lot of detailed content on the blended surface, (d) illustrates minimising avatars, (e) when someone interacts with a blended surface they become spotlighted, (f) when someone talks they become spotlighted and the rest of the avatars are minimised.}}
    \label{fig:Breakouts}
\end{figure}

CDs mostly used the face-to-face blended f-formation for breakouts. A number of CDs explained this was because it was useful to see their partner's reactions and the work surface in the same field of view \textbf{(Figure \ref{fig:Breakouts})}. 

\begin{quote}
\textbf{CD4} \textit{``...I think it's much easier […] for them to actually work on assignments together if the person's in front of them and they can like see the screen together.''}
\end{quote}

Most CDs (6/10) voiced concerns about having \textbf{too many avatars} inside the whiteboard. CD8 suggested two avatars would be the maximum amount to have mirrored in the whiteboard. Whereas CD9 suggested they would prefer only one person ``in the whiteboard" at a time. There were a variety of different suggestions for managing avatars in breakouts. CD1 and CD10 both suggested avatars could be fully \textbf{minimised}. CD10 voiced that they should appear or become \textbf{spotlighted} when they interact with the whiteboard. 
\begin{quote}
\textbf{CD10:} \textit{``...the thought of five different avatars all around the way might be a bit \textbf{distracting}. Maybe, maybe there are options to \textbf{minimise or to compartmentalise} your view of other participants somewhat similar to like a Zoom conversation [...] maybe their \textbf{avatar appears} when they \textbf{interact with the whiteboard}, and maybe when they're not, they're sort of minimised into a, a top kind of toolbar, I guess.''}
\end{quote}

CD8 and CD9 both expressed concerns about using the face-to-face f-formation when there is detailed information on the board. CD9 even suggested having a z-shaped f-formation where the speaking avatar is slightly off the side of the whiteboard so the \textbf{content is less obstructed} \textbf{(Figure \ref{fig:Breakouts})}.
\subsubsection{BCP5: \textsc{Entry zone}: immerse participants in a new environment or prepare them for collaboration}
\begin{figure}
    \centering
    \includegraphics[width=1\linewidth]{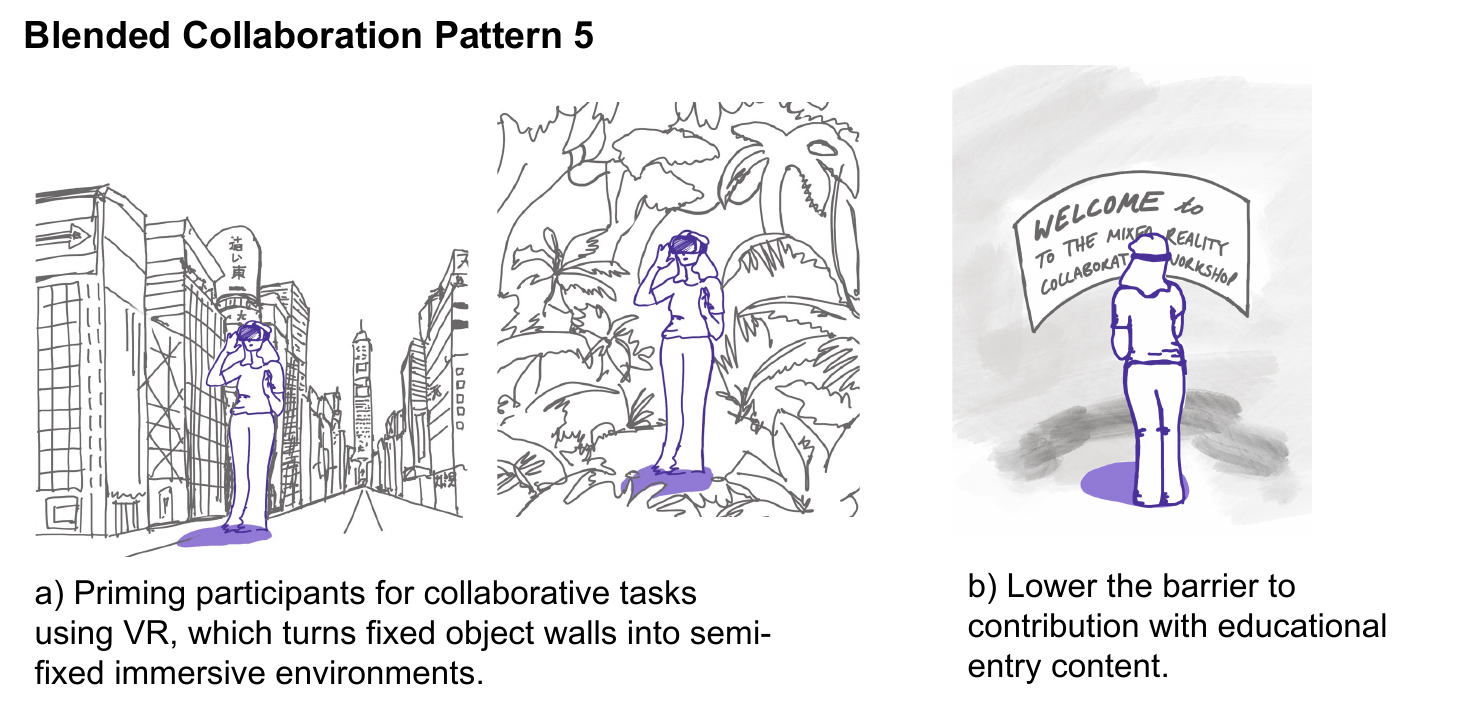}
    \caption{\textbf{DM5) Illustration of interactions suggested for an \textsc{entry zone}, where (a) immerses participants in an inspirational place and (b) teaches them skills that will help them contribute, prior to entering the event}}
    \label{fig:Entry}
\end{figure}

The \textsc{entry zone} supports\textbf{ priming participants }for the collaborative work they will undertake \textbf{(Figure \ref{fig:Entry})}. This zone is flexible depending on the problem being solved. It seeks to \textbf{immerse collaborators} in a new cultural place, which relates to design principle I2, where CDs use place references to influence how collaborators feel, think or act. Other CDs expressed a need to educate and equip participants with practical tools that will help them solve the problem at hand. For example, running a training session when participants enter collaborative MR. 

\subsubsection{BCP6: \textsc{Inspiration and connection zones} identify objects other than surfaces to blend in the MR environment}
\begin{figure}
    \centering
    \includegraphics[width=1\linewidth]{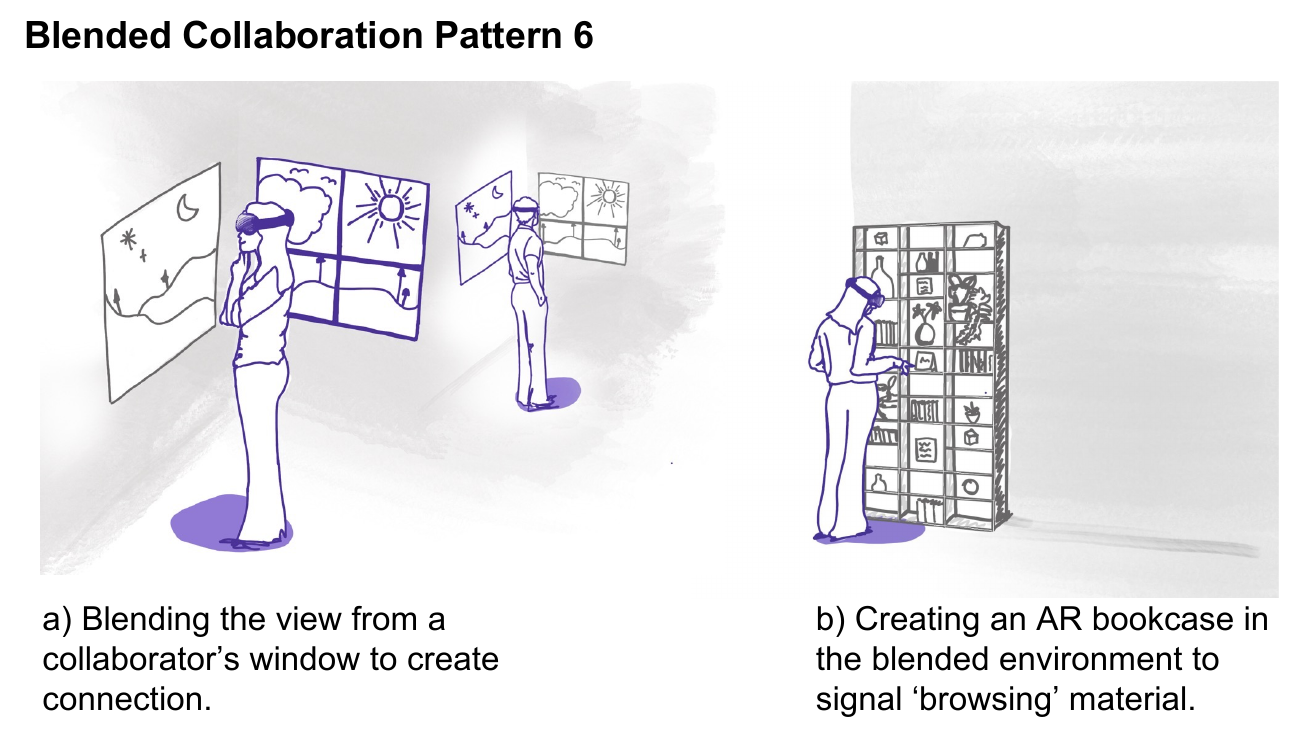}
    \caption{\textbf{BCP6) Illustration of interactions suggested for an \textsc{inspiration and connection zone}, which shows (a) sharing a view from a window and (b) sharing items on a bookcase}}
    \label{fig:InspirationAndConnection}
\end{figure}

The \textsc{inspiration and connection zone} is highly malleable and has the capacity to \textbf{bring physical objects} from the participant's room into the collaborative MR shared environment and display them in different ways \textbf{(Figure \ref{fig:InspirationAndConnection})}. This design choice relates to the design principles I1 and I3, where CDs attempt to influence how participants feel, think or act by placing defined objects in their shared collaborative environment. For example, CDs suggested creating empathy by sharing personal details from a person's home office, such as a \textbf{view} from their \textbf{window} or a \textbf{bookshelf}. However, CD9 expressed that shareable objects might need to be\textbf{ predefined} before entering the blended space for \textbf{privacy} reasons. 

\subsubsection{BCP7: Blended transitions allow movement between different \textsc{collaboration zones} for scaled remote spaces}
\begin{figure}
    \centering
    \includegraphics[width=1\linewidth]{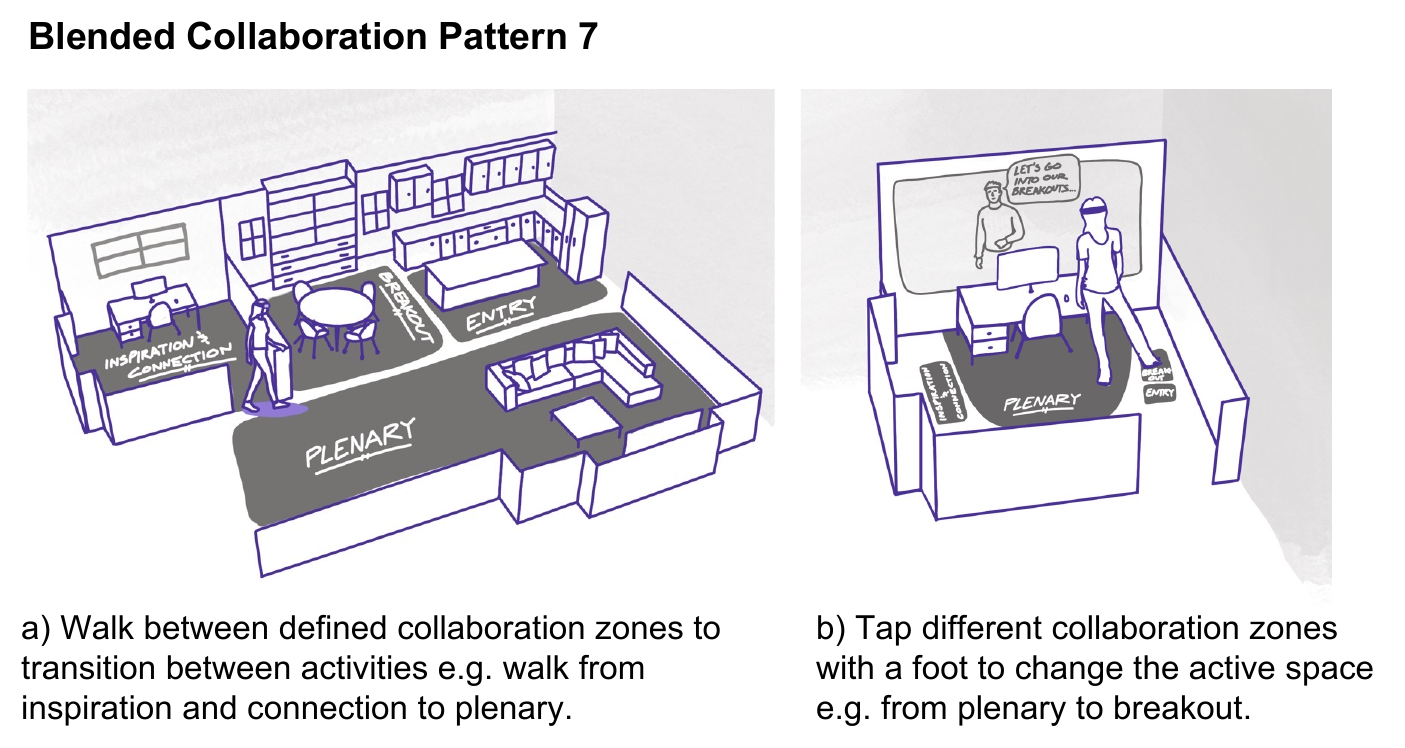}
    \caption{\textbf{BCP7) Illustration of interactions suggested for a single remote user transitioning between different \textsc{collaboration zones}.}}
    \label{fig:Transitions}
\end{figure}

CDs suggested a variety of ways to transition between \textsc{collaboration zones} \textbf{(Figure \ref{fig:Transitions})}. This was done to process, facilitate, and\textbf{ maintain flow} in their scaled MR collaboration spaces, which relates to several categories in the flow state design principles. 

Two of the most relevant transition suggestions were (1) letting participants move between zones using the proxemic layout of fixed and semi-fixed objects in their physical space architecture (although this presents some challenges for a home working space); (2) using the participant's floor to define different zones that the user can \textbf{step into} to shift collaboration areas. 
\begin{quote}
    \textbf{CD1:} \textit{``I imagined that to sit within the plenary as part of like entering the session, you would be asked to \textbf{predefine a small-ish space} that acts as like the \textbf{border between plenary and breakouts} [...] To exit the plenary, they would step outside of that predefined space and that [would then] \textbf{trigger breakouts}.''}
\end{quote}

Others suggested including \textbf{animated transitions} or clicking a button to move, which are both covered in previous research \cite{gronbaek2023partially, fink2022relocations}.  

Three CDs chose to create hybrid collaboration solutions where some participants were co-located, and others joined as single users remotely. A common problem raised was that co-located participants would have their faces covered by the headsets, making it difficult to communicate emotion. 

\subsubsection{BCP8: Role-based point of view enables different proxemic positions or room visibility depending on role and preference}
\begin{figure}
    \centering
    \includegraphics[width=1\linewidth]{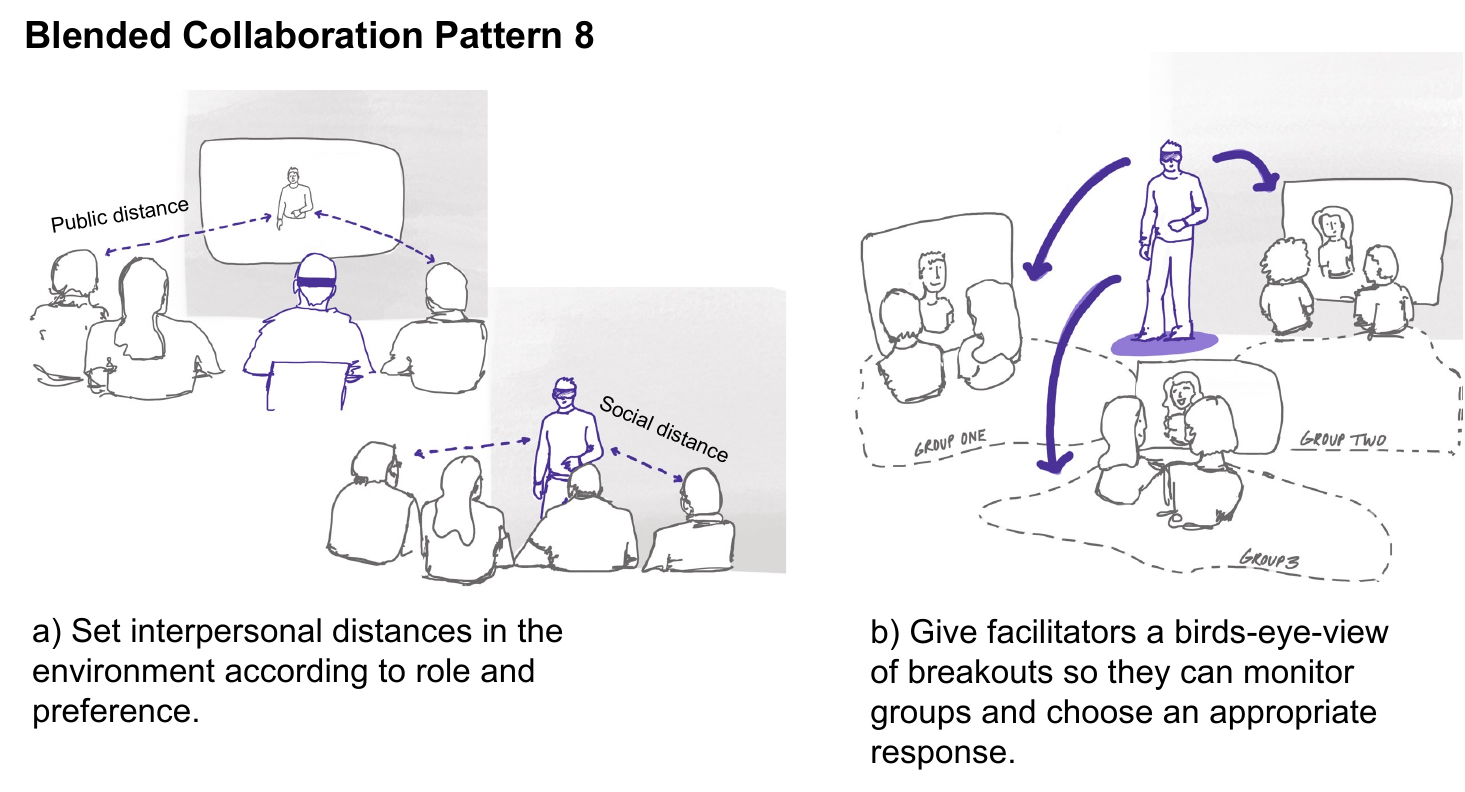}
    \caption{\textbf{BCP8) Illustration of how collaborative MR might display different interpersonal distances and interactions depending on role and/or preference. These ideas were not explicitly prototyped by CDs but rather raised as ideas during the post-prototype discussion. We have created these visuals to help illustrate how different role interactions might work in MR.}}
    \label{fig:roles}
\end{figure}

During the user study, CDs talked about the \textbf{different distances} participants choose to sit, depending on their role and preference \textbf{(Figure \ref{fig:roles})}. This relates to Hall's zones of interpersonal distance \cite{hall1966hidden}, where more introverted participants prefer to sit at a public distance from the speaker and avoid the front row \cite{lawson2007language}. This observation becomes interesting under the light of design principles FS2 and I1. It raises questions about options for interpersonal distance. Should comfort levels be supported such that people can choose how close they feel to the presenter in collaborative MR? Can this option be set differently for the presenter versus the listeners? Data from the CD interviews suggests that, in some situations, participants should be able to\textbf{ choose}, but in others, they want to\textbf{ influence} how participants feel by making them sit closer together.

\begin{quote}
\textbf{CD6:} \textit{``...originally we had the [leaders] on \textbf{highchairs at the front} so that people could see them. And the conversation was really \textbf{awkward}, and it was really kind of \textbf{stifled}. And we changed the [leaders] to be \textbf{sitting on chairs lower} than the [participants], and the conversation just shifted completely.''}
\end{quote}

CDs also expressed a need for facilitators to have a \textbf{different vantage point} of the MR space in order to support transitions and the flow of work. They should be able to \textbf{monitor} how work is progressing across multiple spaces. One CD designed the distributed spaces so the facilitator had a larger room. This was so the facilitator could visualise the different break-out groups in separate spatial areas and walk between these groups, monitoring the progress of participants \textbf{(Figure \ref{fig:roles}, example (b)).} 

\subsection{Applying the design principles}
Applying design principles to CDs' collaborative MR design patterns showed that most patterns are related to \textsc{influence} or \textsc{flow state}, with \textsc{affective response} being the most referenced design principle. \textbf{Figure \ref{fig:DesignModules}} provides a breakdown of BCPs aligned to their corresponding design principles.

\begin{figure}
    \centering
    \includegraphics[width=1\linewidth]{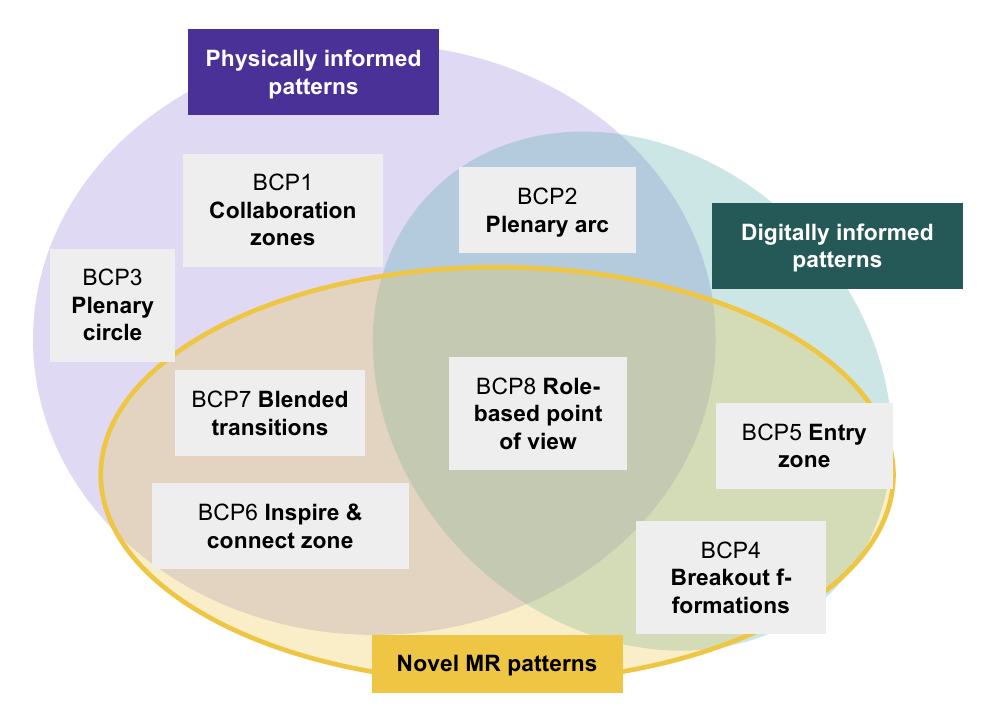}
    \caption{\textbf{Blended collaboration patterns (BCP) for scaled collaboration suggested by CDs mapped to the type of space metaphor that has informed them. These metaphors might be physically informed (e.g. sitting in a circular formation), digitally informed (e.g. minimising speakers), or they could be novel approaches that are made possible by MR (e.g. blended f-formations). The bottom yellow section highlights collaborative interactions that are made possible by MR.}}
    \label{fig:DMMap}
\end{figure}

During the analysis, CDs' prototypes revealed a motif of using physical or digital space metaphors to understand the novel collaborative MR system. \textbf{Figure \ref{fig:DMMap} }visually presents collaborative MR blended collaboration patterns and their alignment with physical or digital space interactions.

\section{Discussion}
In this section, we explore the original research question: \textbf{how might collaboration designers re-imagine distributed mixed reality spaces for large-scale collaboration?} 

CDs offered a number of blended collaboration patterns (BCP) for MR at scale, such as blended f-formations and blended place references. These BCP create a starting point for future research to evaluate scaled interactions in more detail. We recommend researchers explore these BCPs from different user vantage points. This will help to mitigate complexities in trade-offs between visibility and maintaining spatial communication across distributed spaces, to achieve the best possible experience from different user vantage points and roles. Alternately, there is an opportunity to create prototyping tools that can rapidly show how interactions across scaled MR environments will work in practice. 

\subsection{Collaboration designers' blended f-formations illustrate scale trade-offs for future research}
Although MR helps distributed participants collaborate across misaligned spaces \citep{gronbaek2023partially}, these interactions become complicated by overlapping scale issues, such as having multi-users, multi-roles, mixed presence and various asymmetrical surfaces or space architecture. While CDs presented several options for maintaining a collaborative flow state in these scaled situations, their ideas require further work to evaluate these interactions. However, their ideas provide a starting point to pursue Ens et al.'s suggestion that future research explore large-scale group work in MR that better reflects real-world complexities \cite{ens2019revisiting}. 

\begin{figure}
    \centering
    \includegraphics[width=1\linewidth]{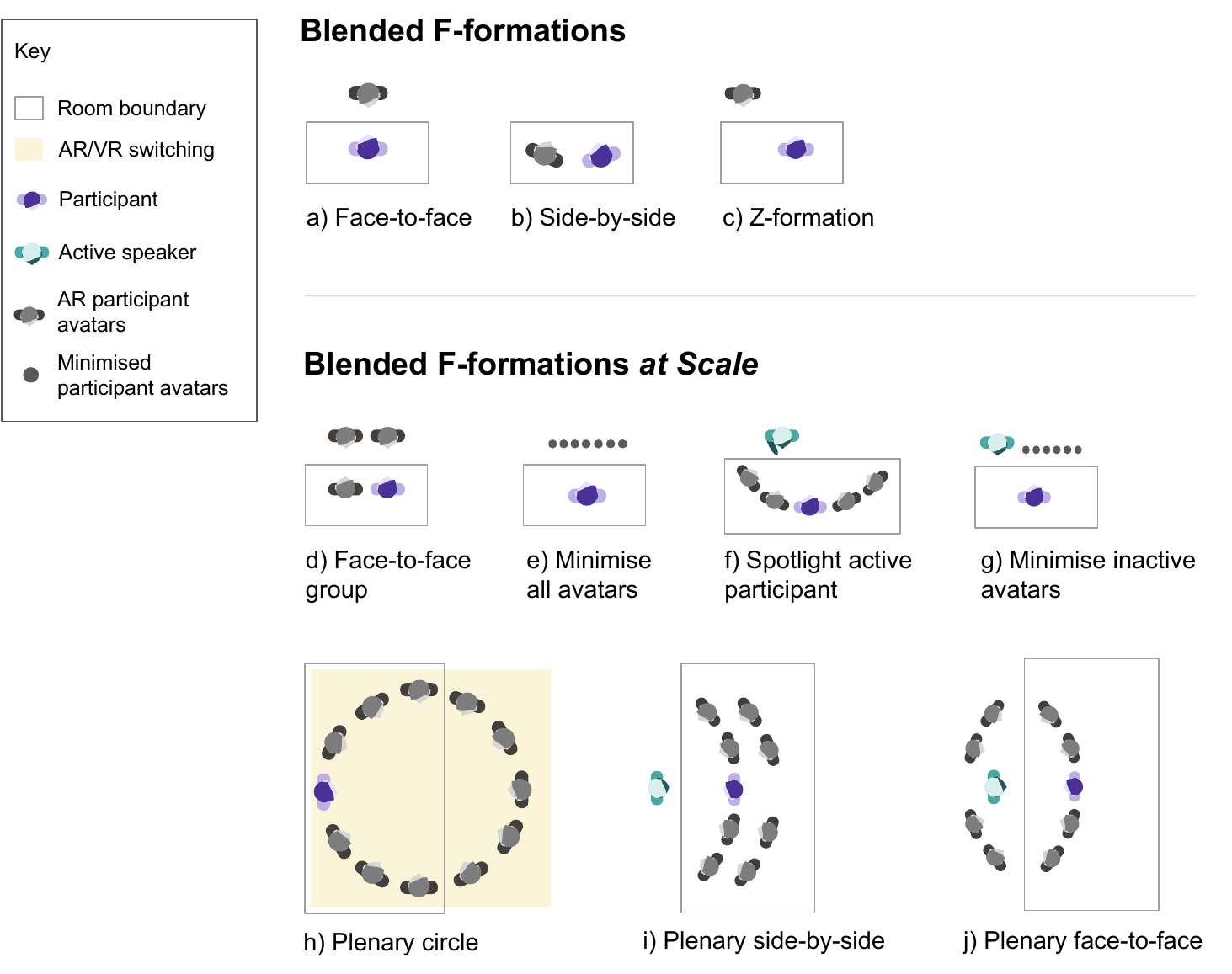}
    \caption{\textbf{(a), (b) and (c) show the different blended f-formations often used in the CDs' prototypes. (d-j) shows a collation of the different blended f-formations and interactions that CDs used to enable scaled collaboration across distributed environments.}}
    \label{fig:PBRf-formations}
\end{figure}

While the blended plenary f-formation suggested by CDs often took the same general shape as Kendon's semi-circular f-formation \citep{marquardt2012cross} \textbf{(Figure \ref{fig:PBRf-formations}, formation i)}, CDs were concerned that scaling beyond five participants would compromise the visibility of content or make smaller spaces feel crowded. For example, imagine fitting fifty avatars into a small home office space. Their concerns align with collaborative \textsc{flow state} design principles FS1, \textsc{remove distractions} and FS2, \textsc{human-centred collaboration}. To mitigate this scaling challenge, CDs suggested creating blended f-formations that leverage both face-to-face and semi-circle formations to extend the room space \textbf{(\textbf{Figure} \ref{fig:PBRf-formations}, formation (j))}. This allows participants to catch \textit{glimpses} of other collaborators through their whiteboard surface and is intended to emulate the ambient feeling of being in a crowd without physically crowding the space \textbf{(Figure \ref{fig:PlenaryArc})}. Using blended f-formations to create space for one-to-many communication is also suggested by He et al.'s MR research \cite{He2020CollaboVR}. However, when we scale to more than four participants in a plenary situation, it is unclear how positioning avatars in blended f-formations will work in practice. Should the system auto-arrange distributed participants sitting in their office chairs into f-formations? If so, will this auto-arrangement break the illusion of co-located alignment, which is important for spatial communication in MR? For example, if the system renders all participants in the centre of the group, then all the avatars would overlap. Instead, each participant might experience being in the `central' position in their own instance while the system renders them in the semi-circle for the presenter and their fellow collaborators. The first-person user might be central in their vantage point but sitting in the second row to the right from another user's first-person vantage point. A higher-fidelity experiment might explore how this scaled asymmetric positioning would work in practice and how these asymmetries affect spatial communication. 

Similarly, CDs suggested solutions to avoid having too many avatars `inside' the whiteboard while working on content in breakouts. Aligned to Interaction Proxemics \cite{mentis2012interaction} theories, some CDs suggested only highlighting the active speaker when they talk or draw on the blended surface. Highlighting the active avatar included making inactive avatars smaller and active ones larger or placing active avatars in the direct field of view \textbf{(Figure \ref{fig:Breakouts}, formation f and g)}. It is difficult to predict how people will respond to this type of interaction that uses the participant's body position to highlight them as an active speaker. Future research might test how collaborative MR handles selectively displaying avatars, exploring the trade-offs between a less distracting environment and being able to view the avatars' body language.

The circular formation \textbf{(Figure \ref{fig:PBRf-formations}, formation h) }was a common suggestion made by CDs as it invites participants to have a more intimate dialogue with each other \cite{coullomb2017collaboration}. Understandably, participants expressed concerns that a headset would cover co-located participants' facial expressions. This problem could be fixed using face display technology similar to Apple's Vision Pro headset. The circle formation may also be challenging for participants joining from a smaller home office, where physical objects might interfere with the avatars' circular formation. Instead, future work might explore creating the ideal activity space for certain formations, such as an unobstructed circle, using a switch between AR/VR modes to remove obstructive physical objects from the MR space.

While the f-formations we observed were similar to Kendon's \textbf{(Figure \ref{fig:PBRf-formations})}\citep{kendon2010spacing}, CDs would create several f-formations to show how focused group work could occur in parallel. Their designs extend beyond previous work that focused on interactions in a single group \cite{kendon1990conducting, marquardt2012cross, marshall2011f-formation, He2020CollaboVR} and zooms out to capture interactions between multiple f-formations. While breakouts are a feature of video conferencing tools such as MS Teams and Zoom, it was interesting to see how CDs facilitated these breakouts spatially using MR. In particular, CDs often designed a facilitator role with a birds-eye view of all the different break-out f-formations \textbf{(figure \ref{fig:roles})}. Having a birds-eye view helps the facilitator assess group progress at a glance and jump from room to room. This can be related back to the collaborative space design principle \textsc{mobile guides}, where the facilitator can assess the room's progress and maintain a collaborative flow state. Future research might focus on this zoomed-out perspective and explore how interactions between multiple f-formations are enabled, both for participants and the facilitation team. 
\subsection{Blended objects go beyond f-formations to support place making}
The ability for spaces to reference other places can influence collaborative behaviour in collaborative MR \cite{sharma2017framework}. CDs inspire curiosity by mimicking how playschools set up their spaces, thereby non-verbally communicating and influencing how participants think, feel, or act \citep{buxton2009mediaspace}. Many of the thematic BCPs synthesised from the results reference other `places' to influence collaborators \citep{harrison1996re, lawson2007language}. These include (1) the \textsc{entry zone}, where collaborators are placed in different VR places to prepare them for collaboration; (2) the \textsc{connection and innovation zone}, which brings in a library bookshelf and other objects from specific places; and (3) f-formations in \textsc{plenary}; where CDs use a theatre-style arc to inspire collective ownership and memory. This blending of places adds to Dourish and Bell's idea that digital remote collaboration tools create spaces layered on spaces \citep{dourish2011divining}; what CDs are trying to create here are \textit{places layered on places}. While place referencing with AR and VR is not new to MR \citep{sharma2017framework, baker2021SchoolsBack}, what is novel is how these place references are blended between distributed physical spaces. For example, CDs suggested blending the participant's view from their window with their collaborator's physical space, showing the time of day or season in their participant's location. Blending meaningful objects from the distributed collaborators' spaces can be explored to build \textit{shared} places for collaboration.

\subsection{Blended f-formations and place references should be designed using a variety of user vantage points}
\begin{figure}
    \centering
    \includegraphics[width=1\linewidth]{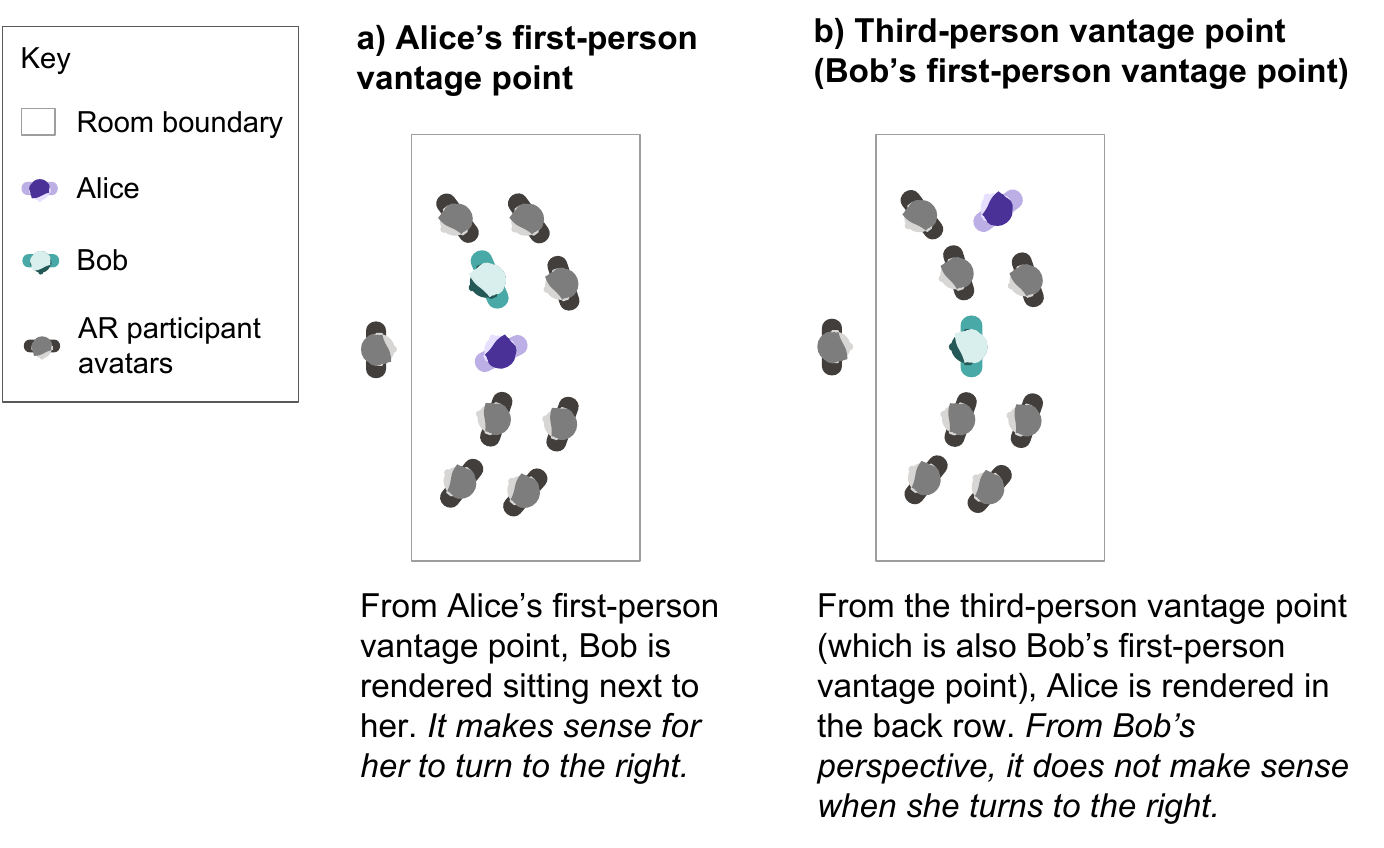}
    \caption{\textbf{Variation in the way avatars are rendered from different vantage points can cause uncanny alignment problems. In this figure, Alice and Bob are rendered in different positions, which changes their spatial communication. Alice looks like she is turning the wrong way from Bob's vantage point.}}
    \label{fig:AliceAndBob}
\end{figure}

To explore scale challenges in distributed collaboration in MR, researchers might examine the CDs' blended f-formations from vantage points other than the first-person. We identified four key vantage points, including (1) the first-person, (2) the third-person, (3) the facilitator's birds-eye view, and (4) the CD's vantage point. Due to scale challenges, each vantage point might view interactions differently, causing miscommunication. For example, from the first-person vantage point in plenary, Alice might be in the front centre row. However, from the third-person vantage point (which can also be thought of as Bob's first-person vantage point), the system may render Alice in the second row to the right. While this variation in avatar positioning can be used to manage the spatial placement of large-scale groups quickly, it requires careful consideration to avoid confusion amongst participants. It would be challenging if Alice turned to talk to the person next to her, thinking it was Bob, when really the system rendered Bob's first-person vantage point so that Alice was not seated next to him, as illustrated in \textbf{figure \ref{fig:AliceAndBob}}. Future work might explore blended f-formations from these different vantage points to understand scale trade-offs between managing avatar visibility, obeying physical boundaries and maintaining just enough what you see is what I see (WYSIWIS) to support communication. 
As the design principles demonstrate, a facilitation and CD vantage point or \textsc{mobile guides} helps ensure collaborators meet their collective objectives. This principle aligns with Goffman's concept of the \textit{performative self} \citep{goffman2016presentation}, which has been adapted by architectural scholars such as Lawson \cite{lawson2007language} to explain how spaces can be designed to facilitate a seamless experiences. For example, in a traditional restaurant setting, diners are seated at the front of the space and the busy kitchen environment is kept hidden ``back-of-stage'' to facilitate a relaxing ``front of stage' dining experience. Connected to this, the principle of \textsc{mobile guides} helps CDs create a performative, simplified voice that guides participants, while back-of-stage processes allow CDs to monitor groups and change the collaborative direction where necessary. As previously discussed, CDs often designed this capability from a birds-eye vantage point to see all the spatial references, such as participant body language, content and layout, all at once. While CDs mostly suggested this view, future work might focus on this facilitator vantage point, examining whether a spatial birds-eye view or a more abstracted two-dimensional format better facilitates back-of-house collaborative processes. 

Alternatively, enabling different vantage points in MR raises ethical questions about asymmetric points of view. While it is important for the facilitation team to have extra abilities to manage collaborative processes, an ability to influence the environment is an important aspect of place-making \citep{harrison1996re}. How much control do participants have to build a place for collaboration versus leaving this up to a facilitator? Does accessing different vantage points challenge facilitator-participant privacy?  It is important for future research in MR collaboration to consider these different roles and ask how space affordances and proxemics might be different from each vantage point.

\subsection{Tools for designing MR interactions for distributed collaboration at scale}
MR needs a design language to visualise how blended f-formations and place references are supported across multiple physical spaces with varying symmetries and user vantage points. Large-scale remote collaboration in MR means the number of spaces and objects defined in the system increases. These spaces might have different symmetries and user vantage points; some may drastically differ in size and alignment, or others may have variations of co-located and remote users. This variation in symmetries and vantage points complicates interaction designs. However, it is difficult to conceptualise how spatial interactions will work across multiple-scale challenges such as multi-users, roles, mixed presence and space asymmetries. For example, if Alice is in a room where her desk is further away from her whiteboard than Bob's, it will take her longer to walk to her desk, breaking the illusion of co-location \cite{gronbaek2023partially}. This asymmetry becomes more complicated with additional layers of scale, such as multiple users or mixed presence. Future research might explore tools to help designers visualise how distributed spaces with multiple users will overlap and prototype interactions within these. For example, miniature models and that might shift the user into different vantage points \cite{herskocitz2022Xspace, pfeuffer2017gaze}, options to add fake avatars or jump in the shoes of different users.

\subsection{Methodological considerations and future work}
This paper deliberately focuses on gathering ideas from collaboration designers (CDs). While this specific group of participants from a particular expertise helps to discover interesting collaborative interactions, it also introduces some limitations.
Firstly, when we used a snowball sampling technique, most of the final participants (9/10) came from one organisation that use the MG Taylor collaboration methodology \citep{mgtaylor}. While sampling from this group helps hone in on large-group collaboration in MR, we note that not all people are familiar with this methodology and do not follow its exact processes during collaboration. However, this approach still holds merit, as many processes aim to improve collaboration. 
Secondly, the prototypes CDs created were biased toward the first-person and CD perspective. Future research should consider how interactions appear from different points of view, including the third-person. This will control for miscommunication due to uncanny alignment situations since MR warps spatial configurations to give an illusion of co-location. 
Thirdly, CDs are not human-computer interaction experts, and their prototypes may be constrained by their limited knowledge of MR interactions. This means that their ideas do not directly consider what is technically feasible and were often coloured by other known digital experiences such as group work on Microsoft Teams or Zoom. However, their limited understanding of MR is also beneficial for the paper's aim to elicit new ideas and inspire areas for future research that may have been overlooked by a researcher constrained by their understanding of what is possible in MR. \rev{The resulting blended collaboration patterns (BCP) should be treated as a starting point for future work and not as an absolute template for creating all future MR collaboration tools.} This limitation offers an opportunity for human-computer interaction (HCI) experts and AR/VR designers and developers to evaluate these BCP from their HCI domain of expertise in future studies. 
Finally, in the study, we used low-fidelity analogue prototyping tools, which might have constrained the design space during the solution exploration. Thus, future work could incorporate a broader range of prototyping tools within AR/VR for more high-fidelity prototyping, such as \rev{\textit{TiltBrush}\footnote{https://www.tiltbrush.com/}, \textit{ShapesXR}\footnote{https://www.shapesxr.com/} or \textit{Roblox}\footnote{https://create.roblox.com/?}}. While different headsets will have some variance in the overall user experience, the BCP have been designed to be device agnostic. Any device that uses video see-through technology to enable seamless AR/VR switching, such as the Meta Quest Pro or Apple Vision Pro, should be able to support testing the BCPs with high-fidelity prototypes.

\section{Conclusion}
Our research explores distributed MR for collaboration at scale, shifting focus from dyads to complex scale interactions with multiple users, roles, mixed presence and space size. Through workshops, we let collaboration designers (CDs) play a central role in reimagining distributed MR spaces and identified three core design principles: maintaining flow-state, influencing the way collaborators think, feel or act, and responding to activities in the space. The synthesised blended collaboration patterns (BCP) offer a starting point for future research to study how blended facing formations (f-formations) and blended place references will support scaled collaboration in MR. The trade-offs between spatial communication and visibility created by these interactions should be explored from vantage points other than the first-person; viewing interactions from the third-person, facilitator birds-eye and collaboration designer vantage points. However, prototyping from these vantage points in MR can be challenging. Future work might explore how different levels of fidelity and visual representation can be used to demonstrate the trade-offs between spatial alignment and visibility during MR collaboration at scale. By looking from a CD vantage point, this research defines a series of opportunities that will help to build MR systems that support large-scale collaboration.

\begin{acks}
We thank all the MG Taylor practitioners who provided invaluable insight into how they design collaborative spaces. This study would not have been possible without the life work of Matt and Gail Taylor. We are grateful for their deep insight into collaborative space design, which has been passed between generations of practitioners.
This research was funded by the Independent Research Fund Denmark (DFF) (grant ID: 10.46540/3104-00008B), Aarhus University Research Foundation (AUFF) (grant ID: AUFF-F2021-7-2), and the Pioneer Centre for AI, DNRF grant number P1.
\end{acks}

\bibliographystyle{ACM-Reference-Format}
\bibliography{software}

\appendix

\section{Low-Fidelity Prototype Study Design and Participant Sample}

\subsection{Questions given to participants to prompt ideas for their low-fidelity prototype}

Similar to Rajaram et al.'s study \citep{rajaram2023eliciting}, participants were given questions to prompt their ideas. These questions were informed by the collaboration patterns Reilly et al. used to co-design their media space \citep{reilly2010space} as follows:
\begin{enumerate}
     \item How does the plenary work? What does it look like for people who join from their own space? Should people be able to sit ‘beside’ each other?
     \item How do digital artefacts blend with the participant’s space? What is displayed, and how is it displayed?
     \item  How would breakout areas work?
     \item What does an entry experience look like?
 \end{enumerate}

Participants were asked to "brainstorm ways you would adapt the mixed reality technology shown, to design and facilitate a collaborative event". The questions above were given as a guide only and the interviewer told participants they were free to deviate from these as they saw fit. 

\subsection{A summary of each participant's years of collaboration experience and familiarity with mixed reality (MR) tools}

\begin{table}
\caption{Collaboration designer (CD) participants in the three-part expert workshop: a break down of their years of collaboration experience and familiarity with MR tools.}
 \label{tab: workshopParticipants}
\begin{tabular}{p{0.5cm} p{1cm} p{1.5cm} p{2cm}}
 \toprule
 CD ID&Pronouns&Years of collaboration experience&Experience using MR\\
 \midrule
  CD1& she/her& 2& I have used it a handful of times\\
  CD2&  he/him& 12& I have used it a handful of times\\
  CD3& she/her& 2.5& I have never used it\\
  CD4& she/her& 7& I have used it once or twice\\
  CD5& he/him& several (prefer not to say)& I have used it once or twice\\
  CD6& she/her& 5& I have used it a handful of times\\
  CD7& prefer not to say& 7& I have used it a handful of times\\
  CD8& he/him& 11& I have used it a handful of times\\
  CD9& she/her& 9& I have used it once or twice\\
  CD10& he/him& 7& I have never used it\\
  \bottomrule
\end{tabular}
\end{table}

\begin{table}
\caption{Collaboration designer (CD) participants who validated the design principles: a break down of their years of collaboration experience.}
 \label{tab:validationParticipants}
\begin{tabular}{p{0.5cm} p{1cm} p{2.5cm}}
 \toprule
 CD ID&Pronouns&Years of collaboration experience\\
 \midrule
  CD1& he/him& 12\\
  CD2& he/him& 6\\
  CD3& she/her& 2.5\\
  CD4& he/him& several (prefer not to say)\\
  CD5& he/him& 11\\
  CD6& she/her& 9\\
  \bottomrule
\end{tabular}
\end{table}

\end{document}